\documentclass[
reprint,
superscriptaddress,
 amsmath,amssymb,
 aps,
prb,
floatfix
]{revtex4-2}
\usepackage{graphicx}
\usepackage{dcolumn}
\usepackage{amsmath, amssymb}
\usepackage{dsfont}
\usepackage{comment}
\usepackage{bm}
\usepackage[colorlinks=true,citecolor=green,linkcolor=red,urlcolor=blue]{hyperref}

\usepackage{tikz}
\usetikzlibrary{decorations.pathreplacing}
\usetikzlibrary{patterns}

\usepackage{xstring}
\usepackage{physics}

\newcommand{\dc}[0]{\mathcal{Z}(\mathcal{C})}

\usetikzlibrary{calc, decorations.markings}
\usetikzlibrary {shapes.geometric}
\usetikzlibrary{arrows.meta}
\colorlet{D1}{black}
\colorlet{D2}{blue}

\usepackage[caption=false]{subfig}
\usepackage{cleveref}
\usepackage{color}

\newcommand{\be}{\begin{equation}}
\newcommand{\ee}{\end{equation}}
\newcommand{\beqn}{\begin{eqnarray}}
\newcommand{\eeqn}{\end{eqnarray}}

\usepackage[normalem]{ulem}

\begin{document}

\preprint{APS/123-QED}

\title{Extended string-net model with all anyons at finite temperature}

\author{Andr\'e Soares}
\email{
andre-octavio.pais-ferreira-dossantos-soares@universite-paris-saclay.fr}
\affiliation{Sorbonne Universit\'e, CNRS, Laboratoire de Physique Th\'eorique de la Mati\`ere Condens\'ee, LPTMC, F-75005 Paris, France}

\author{Anna Ritz-Zwilling}
\email{anna.ritz\_zwilling@sorbonne-universite.fr }
\affiliation{Sorbonne Universit\'e, CNRS, Laboratoire de Physique Th\'eorique de la Mati\`ere Condens\'ee, LPTMC, F-75005 Paris, France}

\author{Jean-No\"el Fuchs}
\email{jean-noel.fuchs@sorbonne-universite.fr}
\affiliation{Sorbonne Universit\'e, CNRS, Laboratoire de Physique Th\'eorique de la Mati\`ere Condens\'ee, LPTMC, F-75005 Paris, France}

\begin{abstract}
The string-net model describes a vast  family  of  topological  orders  in  two  spatial  dimensions. Here, we consider the effect of thermal fluctuations on these topological phases. In the original string-net model, the description of charge (vertex) excitations can be problematic. Therefore, in order to describe all anyon excitations, we study an extended model [Y. Hu, N. Geer, Y.-S. Wu, \emph{Phys. Rev. B} $\bold{97}$, 195154 (2018)]. Building on recent methods, we compute the spectral degeneracies of excited states and obtain the exact partition function. In the thermodynamic limit, the latter is dominated by the trivial (vacuum) anyon, so that topological order is destroyed at any non-zero temperature. In contrast, in a finite-size system, order survives up to a finite temperature, revealing a non-trivial scaling between temperature and size similar to that of the one-dimensional classical Ising model.  We confirm this scaling by computing the thermal average of several observables such as Wegner-Wilson loops and topological mutual information.
\end{abstract}

\maketitle

\section{\label{sec:Introduction} Introduction}

Topological orders describe a class of quantum phases of matter at zero temperature that lie outside the scope of the symmetry-breaking paradigm \cite{landau2013statistical}. In two dimensions and at zero temperature, these gapped phases are characterized by the robustness of their topological ground-state degeneracy against small local perturbations \cite{QInfoMeetsQMatter} and their point-like quasiparticles, called anyons, that display exotic exchange statistics \cite{FirstAnyons1,FirstAnyons2}. They can be divided into two main classes: \emph{chiral} topological order, which hosts gapless chiral edge states and breaks time-reversal symmetry, and \emph{achiral} topological order, which has gappable edges and is (often) time-reversal symmetric. Although systems in the former class are relevant experimentally, since they notably include the fractional quantum Hall effect~\cite{Tsui_1982},  achiral topological orders are also interesting because they can be realized with simple lattice toy models~\cite{Wen_2017}, which permit artificial realisations, e.g. with superconducting qubits~\cite{Song_2018,Satzinger_2021,Google2021,Xu2024}. The latter include the Kitaev quantum double model (e.g., the famous toric code)~\cite{Kitaev_2003,Kitaev2006}, Levin and Wen string-net model~\cite{Levin_2005} and their generalizations~\cite{SNThorough,LanWen,Lin_2014,Hahn_2020,Lake2016}.

The abstract algebraic structure of $(2+1)$-D topological order is provided by fusion category theory, which is crucial in understanding the topological properties of emergent excitations (anyons)~\cite{Kitaev2006}. From a category-theoretical perspective, string-net models take as input an unitary fusion category (UFC) $\mathcal{C}$~\cite{etingof2015tensor} that describes the microscopic degrees of freedom and, by embedding it on a lattice, they realize a topological order characterized by the Drinfeld center $\dc$, which is a unitary modular tensor category (UMTC)~\cite{LanWen, Kitaev2012, Kong2012}. Although the Drinfeld center encodes several properties of the emergent anyons (e.g. exchange statistics and topological spin), the lattice model uncovers additional local degrees of freedom, or subtypes, of these anyons, that are not preserved under local perturbations~\cite{Scipost_DiTc,K_m_r_2017} and require more information than that contained in $\dc$.

The nature and impact of these added degrees of freedom on the energy spectrum are still uncharted. Recently, all the energy level degeneracies of the Levin-Wen string-net model were computed analytically, which allowed for the computation of the partition function and several observables at finite temperature~\cite{Ritz_Zwilling_2024,Ritz-Zwilling2024_2}. A main conclusion of that work was that topological order is destroyed at any finite temperature in the thermodynamic limit, due to the entropic proliferation of thermally-generated anyons. This is in line with previous results which have shown the thermal fragility, or destruction, of topological order at any finite temperature in the toric code~\cite{Castelnovo2007,Nussinov2008}, and more generally in any two-dimensional Hamiltonian that is a sum of local commuting projectors \cite{Hastings2011}. Nonetheless, in finite-size systems, these studies have shown that there is a finite-temperature below which topological order persists.

It is well-known that creating pure vertex excitations can be problematic in the Levin-Wen string-net model~\cite{Hu_2018}, although this can be achieved when the input data corresponds to a group, as is the case of the toric code topological order~\cite{Levin_2005}. It is therefore usually studied in the absence of branching rules violations (i.e. in the ``charge-free sector'', without vertex excitations), a drawback that prohibits the study of some anyon excitations. To circumvent this, Hu, Geer and Wu~\cite{Hu_2018} proposed an extension of the string-net model, by attaching a tail for each vertex of the original lattice, effectively simulating vertex excitations while still respecting the branching rules at every vertex.

In this paper, we study the finite-temperature properties of such an extended string-net model displaying all anyons. Inspired by~\cite{Hu_2018}, we consider a model on the honeycomb lattice where a single tail is added to each plaquette of the lattice, and establish a connection between local operators acting on each plaquette and the tube algebra. From this tube algebra, we show that all anyons can be realized as elementary excitations and obtain their internal multiplicity (or number of subtypes) from which the computation of level degeneracies and partition function follows directly.

These results can be generalized to a family of extended models where plaquettes are allowed to contain an arbitrary number $k$ of tails. The corresponding tube algebras, employed to describe the emergent anyons, must also be extended to allow their tubes to have additional tails. These multi-tails tube algebras have been studied previously~\cite{LanWen, MapMorita}, but their importance is here highlighted because they describe the emergent anyon subtypes, which depend on the number of tails in each plaquette. This procedure simultaneously reveals the fundamental role of the one-tail model: the anyons arising in all these extended models with $k\geq 1$ tails are still described by $\dc$, and their internal multiplicities, computed from the generalized tube algebras, can be determined from the simpler, one-tail tube algebra.

The article is organized as follows. In Sec.~\ref{sec:SNModels}, we introduce the extended string-net models. In Sec.~\ref{sec:TubeAlgebras}, we study the corresponding tube algebras that are needed to construct the Drinfeld center of a given input category. The main results are presented over the next two sections. In Sec.~\ref{sec:Degeneracies}, we compute the spectral degeneracies and obtain the partition function. From these results, in Sec.~\ref{sec:finitesize} we detail how topological orders survive in finite-size systems by computing several finite-temperature observables (partition function, topological projectors, Wegner-Wilson loops and topological mutual information). In the following Sec.~\ref{sec:Examples}, we illustrate these general results with a simple but non-trivial example (a non-Abelian input category called Fibonacci), before concluding in Sec.~\ref{sec:Conclusion}. In two Appendices, we give details on an important subset of anyons $\mathcal{N}$ (App.~\ref{appendix:SpecialSubset}) and on the decomposition of the tube algebra (App.~\ref{appendix:TAdecomp}).

\section{\label{sec:SNModels} String-net models}

In this section, we introduce the relevant models for which the partition function will be determined. First, we review the main ideas behind the original string-net model, following the construction of Levin and Wen (LW)~\cite{Levin_2005}. This model realizes in principle all anyons expected from the Drinfeld center of the input category, but is rarely studied in the full Hilbert space because a vertex excitation automatically excites also the neighboring plaquettes. We therefore consider next a simpler model, here called $LW_0$, by restricting the total Hilbert space to the ``charge-free sector'', meaning that vertex excitations are not allowed. As a consequence, this model does not host all anyons, but only a subset of anyons called fluxons. We finally focus on the extended string-net models~\cite{Hu_2018}, which are slight generalizations of the $LW_0$ model obtained by adding additional degrees of freedom to the lattice (tails). These models achieve a more unified framework to study excited states as they include all anyons of the Drinfeld center, and not just fluxons.

\subsection{Original string-net construction}\label{sec:LWmodel}

String-net models are defined by some input data, corresponding to a UFC $\mathcal{C}$ that describes the local degrees of freedom that live on the edges of a (directed)  trivalent graph embedded on a closed and orientable two-dimensional manifold.

The fusion category $\mathcal{C}$ comprises simple objects (strings) $a,b,c,\dots$, corresponding quantum dimensions $d_a,d_b,d_c, \dots\geq 1$, fusion rules on these objects represented by $a\times b = \sum_{c} N_{ab}^c c$ with $N^c_{ab}$ non-negative integers, and an isomorphism $F:(a\times b)\times c \rightarrow a \times (b \times c)$ that solves the pentagon equation (for an introduction to UFCs, see e.g.~\cite{Simon2023}).  If $|\mathcal{C}|$ denotes the number of simple objects in the UFC $\mathcal{C}$, then the fusion symbols $N_{ab}^c$ can be interpreted as $|\mathcal{C}|\times|\mathcal{C}|$ matrices $N_a$ for each string $a\in \mathcal{C}$, with elements $[N_a]_{b,c}=N_{ab}^{\bar{c}}$ (rows indexed by $b$ and columns by $c$). There is a unique identity object $1$ (also called vacuum string) that fuses trivially with all other elements via $a\times 1 = 1 \times a = a$, such that $N_{1}=\mathds{1}$ is the identity matrix. In addition, each element $a$ has a corresponding dual $\bar{a}$ such that $N_{a\bar{a}}^1 = 1$, $d_a = d_{\bar{a}}$ and $1\equiv \bar{1}$.  Throughout the article, the discussion is restricted to multiplicity-free fusion categories that satisfy $N^c_{ab}\leq 1$, although the results could easily be generalized.

\begin{figure}[t!]
    \subfloat[\label{fig:BasisState}]{ \resizebox{3.5cm}{!}{\rotatebox{70}{
\begin{tikzpicture}

\begin{scope}[line cap=round,decoration={
	markings,
	mark=at position 0.7 with {\arrow{stealth}}},
        line width=0.30mm]

    \coordinate (A1) at (-0.6,0.9);

    \coordinate (A2) at (0.2,0.53);

    \coordinate (A3) at (-0.1823,1.6218);

    \coordinate (A4) at (-2.123213,2.1234);

    \coordinate (A5) at (-1,-0.233);

    \coordinate (A6) at (0.2823,-0.6218);

    \coordinate (A7) at (-1.723213,1.165334);

    \coordinate (A8) at (0.992,0.58233);

    \coordinate (A9) at (1.5823,-0.018);

    \coordinate (A10) at (-1.6234, -0.618);

    \coordinate (A11) at (-2.6234,0.1818);

    \draw[D1, postaction={decorate}] (A1) -- (A2);
    \draw[D1, postaction={decorate}, line width=0.45mm] (A2) -- (A3);

    \draw[D1, postaction={decorate}] (A2) -- (A8);
    
    \draw[D1, postaction={decorate}] (A7) -- (A4);
    \draw[D1, postaction={decorate}] (A3) -- (A4);

     \draw[D1, postaction={decorate}] (A7) -- (A1);

    \draw[D1, postaction={decorate}] (A1) -- (A5);
    \draw[D1, postaction={decorate}] (A8) -- (A6);
    \draw[D1, postaction={decorate}] (A8) -- (A9);

    \draw[D1, postaction={decorate}] (A5) -- (A10);

    \draw[D1, postaction={decorate}] (A5) -- (A6);

    \draw[D1, postaction={decorate}] (A11) -- (A7);
    \draw[D1, postaction={decorate}] (A10) -- (A11);

    \draw[D1, postaction={decorate}] (A4) --++ (210:0.7);
    \draw[D1, loosely dotted,line width=0.20mm] ($(A4) + (210:0.7)$) --++ (210:0.7);

    \draw[D1, postaction={decorate}] (A3) --++ (3.4:0.7);
    \draw[D1, loosely dotted,line width=0.20mm] ($(A3) + (3.4:0.7)$) --++ (3.4:0.7);

    \draw[D1, postaction={decorate}] (A6) --++ (330:0.7);
    \draw[D1, loosely dotted,line width=0.20mm] ($(A6) + (330:0.7)$) --++ (330:0.7);

    \draw[D1, postaction={decorate}] (A9) --++ (100:0.7);
    \draw[D1, loosely dotted,line width=0.20mm] ($(A9) + (100:0.7)$) --++ (100:0.7);

    \draw[D1, postaction={decorate}] (A9) --++ (290:0.7);
    \draw[D1, loosely dotted,line width=0.20mm] ($(A9) + (290:0.7)$) --++ (290:0.7);

    \draw[D1, postaction={decorate}] (A10) --++ (170:0.7);
    \draw[D1, loosely dotted,line width=0.20mm] ($(A10) + (170:0.7)$) --++ (170:0.7);

    \draw[D1, postaction={decorate}] (A11) --++ (150:0.7);
    \draw[D1, loosely dotted,line width=0.20mm] ($(A11) + (150:0.7)$) --++ (150:0.7);
    
    \node[rotate=-70] (A1) at ($0.5*(A1) + 0.5*(A2) + (0.5,0.5)$) {$\boldsymbol{a}$};

\end{scope}

\end{tikzpicture}
}}}\hfill%
    \subfloat[\label{fig:BasisState_Reversed}]{ \resizebox{3.5cm}{!}{\rotatebox{70}{
\begin{tikzpicture}

\begin{scope}[line cap=round,decoration={
	markings,
	mark=at position 0.7 with {\arrow{stealth}}},
        line width=0.30mm]

    \coordinate (A1) at (-0.6,0.9);

    \coordinate (A2) at (0.2,0.53);

    \coordinate (A3) at (-0.1823,1.6218);

    \coordinate (A4) at (-2.123213,2.1234);

    \coordinate (A5) at (-1,-0.233);

    \coordinate (A6) at (0.2823,-0.6218);

    \coordinate (A7) at (-1.723213,1.165334);

    \coordinate (A8) at (0.992,0.58233);

    \coordinate (A9) at (1.5823,-0.018);

    \coordinate (A10) at (-1.6234, -0.618);

    \coordinate (A11) at (-2.6234,0.1818);

    \draw[D1, postaction={decorate}] (A2) -- (A1);
    \draw[D1, postaction={decorate}, line width=0.45mm] (A3) -- (A2);

    \draw[D1, postaction={decorate}] (A2) -- (A8);
    
    \draw[D1, postaction={decorate}] (A7) -- (A4);
    \draw[D1, postaction={decorate}] (A3) -- (A4);

     \draw[D1, postaction={decorate}] (A7) -- (A1);

    \draw[D1, postaction={decorate}] (A1) -- (A5);
    \draw[D1, postaction={decorate}] (A8) -- (A6);
    \draw[D1, postaction={decorate}] (A8) -- (A9);

    \draw[D1, postaction={decorate}] (A5) -- (A10);

    \draw[D1, postaction={decorate}] (A5) -- (A6);

    \draw[D1, postaction={decorate}] (A11) -- (A7);
    \draw[D1, postaction={decorate}] (A10) -- (A11);

    \draw[D1, postaction={decorate}] (A4) --++ (210:0.7);
    \draw[D1, loosely dotted,line width=0.20mm] ($(A4) + (210:0.7)$) --++ (210:0.7);

    \draw[D1, postaction={decorate}] (A3) --++ (3.4:0.7);
    \draw[D1, loosely dotted,line width=0.20mm] ($(A3) + (3.4:0.7)$) --++ (3.4:0.7);

    \draw[D1, postaction={decorate}] (A6) --++ (330:0.7);
    \draw[D1, loosely dotted,line width=0.20mm] ($(A6) + (330:0.7)$) --++ (330:0.7);

    \draw[D1, postaction={decorate}] (A9) --++ (100:0.7);
    \draw[D1, loosely dotted,line width=0.20mm] ($(A9) + (100:0.7)$) --++ (100:0.7);

    \draw[D1, postaction={decorate}] (A9) --++ (290:0.7);
    \draw[D1, loosely dotted,line width=0.20mm] ($(A9) + (290:0.7)$) --++ (290:0.7);

    \draw[D1, postaction={decorate}] (A10) --++ (170:0.7);
    \draw[D1, loosely dotted,line width=0.20mm] ($(A10) + (170:0.7)$) --++ (170:0.7);

    \draw[D1, postaction={decorate}] (A11) --++ (150:0.7);
    \draw[D1, loosely dotted,line width=0.20mm] ($(A11) + (150:0.7)$) --++ (150:0.7);
    
    \node[rotate=-70] (A1) at ($0.5*(A1) + 0.5*(A2) + (0.5,0.5)$) {$\bar{\boldsymbol{a}}$};

\end{scope}

\end{tikzpicture}
}}}
\caption{Part of an oriented trivalent graph. A basis state for the Hilbert space is constructed by labelling each edge with an irreducible object $a \in \mathcal{C}$. The configurations (a) and (b) are regarded as the same state, with the direction of one edge reversed and the corresponding label $a$ conjugated to $\bar{a}$.}
\label{fig:ExampleState}
\end{figure}
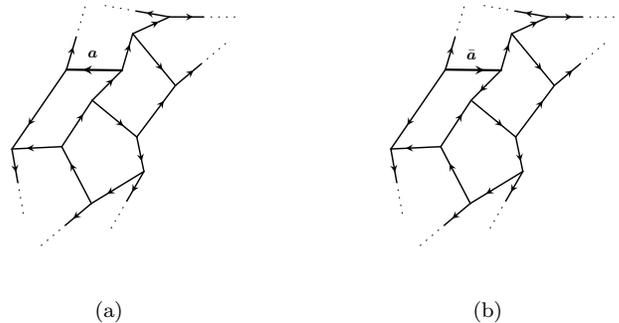

A basis state for the Hilbert space can be constructed by assigning a string from the input UFC to each edge of the trivalent graph, where the degrees of freedom are located. The direction of an edge $a$ can be reversed by assigning it the dual label $\bar{a}$ (see Fig. \ref{fig:ExampleState}).

The Hamiltonian is given by a sum of local projectors of two types:
\begin{equation}
    H = -J_p \sum_p B_p - J_v \sum_v Q_v.
    \label{eq:FullHamiltonian}
\end{equation}
Here, the sums are carried over the plaquettes and vertices of the graph, respectively. A vertex projector $Q_v$ assigns an energy cost to states that violate the fusion rules at $v$. It acts locally on the edges connected to the vertex according to:
\begin{equation}
     Q_v \ket{\resizebox{2cm}{!}{\begin{tikzpicture}[thick,decoration={
		markings,
		mark=at position 0.6 with {\arrow{stealth}}},
		baseline={([yshift=-.5ex]current bounding box.center)},
		line cap=round
	] 
	
\begin{scope}[black] 
    \draw[D1, postaction={decorate}] (0,0) -- (0,0.8);  
    \draw[D1, postaction={decorate}] (-45:0.8) -- (0,0);
    \draw[D1, postaction={decorate}] (-135:0.8) -- (0,0);
	
    \node[D1] at ($(-135:0.8) - (0.15,0.15)$) {$a$};
    \node[D1] at ($(-45:0.8) - (- 0.15,0.15)$) {$b$};
    \node[D1] at ($(0,0.8) + (0.0,0.15)$) {$c$};
	
\end{scope}
\end{tikzpicture}}} = \delta_{a,b,c} \ket{\resizebox{2cm}{!}{\begin{tikzpicture}[thick,decoration={
		markings,
		mark=at position 0.6 with {\arrow{stealth}}},
		baseline={([yshift=-.5ex]current bounding box.center)},
		line cap=round
	] 
	
\begin{scope}[black] 
    \draw[D1, postaction={decorate}] (0,0) -- (0,0.8);  
    \draw[D1, postaction={decorate}] (-45:0.8) -- (0,0);
    \draw[D1, postaction={decorate}] (-135:0.8) -- (0,0);
	
    \node[D1] at ($(-135:0.8) - (0.15,0.15)$) {$a$};
    \node[D1] at ($(-45:0.8) - (- 0.15,0.15)$) {$b$};
    \node[D1] at ($(0,0.8) + (0.0,0.15)$) {$c$};
	
\end{scope}
\end{tikzpicture}}},
\end{equation}
where $\delta_{a,b,c}=1$ if $N_{ab}^c\geq 1$ and 0 otherwise. The $B_p$ projectors have a more complex structure and the interested reader is referred to \cite{SNThorough,Hahn_2020} for details on their definition. In essence, they are defined as:
\begin{equation}
B_p = \frac{1}{\mathcal{D}_{\mathcal{C}}^2}\sum_s d_s B_p^s,
\end{equation}
with $\mathcal{D}_{\mathcal{C}} = \sqrt{ \sum_{s\in \mathcal{C}} d_s^2}$ the total quantum dimension of $\mathcal{C}$. Each operator $B_p^s$ has a straightforward diagrammatic interpretation: it inserts a string $s$ type loop inside the plaquette $p$ and fuses it onto the lattice using the $F$-symbols:
\begin{equation}
     B_p^s \ket{\resizebox{2cm}{!}{\begin{tikzpicture}[thick,
		baseline={([yshift=-.5ex]current bounding box.center)},
		scale=.7
	]

\begin{scope}[line cap=round,decoration={
	markings,
	mark=at position 0.6 with {\arrow{stealth}}}]
	
	\draw[D1, postaction={decorate}] (0,0) -- (30:1.5);
	\draw[D1, postaction={decorate}] (0,0) -- (150:.75);
	\draw[D1, postaction={decorate}] (0,-1.5) -- (0,0);
	\draw[D1, postaction={decorate}] ({cos(30)*1.5*2},-1.5) -- ({cos(30)*1.5*2},0);
	
	\draw[D1, postaction={decorate}] ({cos(30)*3},0) -- ++(30:.75);
	\draw[D1, postaction={decorate}] ({cos(30)*3},0) -- ++(150:1.5);
	\draw[D1, postaction={decorate}] ({cos(30)*1.5}, {-1.5-1.5*sin(30)}) -- ++(150:1.5);
	\draw[D1, postaction={decorate}] ({cos(30)*1.5}, {-1.5-1.5*sin(30)}) -- ++(30:1.5);
	\draw[D1, postaction={decorate}] ({2.5*cos(30)*1.5}, {-1.5-.75*sin(30)}) -- ++(150:.75);
	\draw[D1, postaction={decorate}] ({-cos(30)*.75}, {-1.5-.75*sin(30)}) -- ++(30:.75);
	\draw[D1, postaction={decorate}] ({cos(30)*1.5}, {sin(30)*1.5}) -- ++(0,.75);
	\draw[D1, postaction={decorate}] ({cos(30)*1.5}, {-1.5-2*sin(30)*1.5}) -- ++(0,.75);
\end{scope}

\begin{scope}[line cap=round,decoration={
	markings,
	mark=at position 0.5 with {\arrow{stealth}}}]
\end{scope}
 
\end{tikzpicture}}} = \ket{\resizebox{2cm}{!}{\begin{tikzpicture}[thick,
		baseline={([yshift=-.5ex]current bounding box.center)},
		scale=.9
	]

\begin{scope}[line cap=round,decoration={
	markings,
	mark=at position 0.6 with {\arrow{stealth}}}]
	
	\draw[D1, postaction={decorate}] (0,0) -- (30:1.5);
	\draw[D1, postaction={decorate}] (0,0) -- (150:.75);
	\draw[D1, postaction={decorate}] (0,-1.5) -- (0,0);
	\draw[D1, postaction={decorate}] ({cos(30)*1.5*2},-1.5) -- ({cos(30)*1.5*2},0);
	
	\draw[D1, postaction={decorate}] ({cos(30)*3},0) -- ++(30:.75);
	\draw[D1, postaction={decorate}] ({cos(30)*3},0) -- ++(150:1.5);
	\draw[D1, postaction={decorate}] ({cos(30)*1.5}, {-1.5-1.5*sin(30)}) -- ++(150:1.5);
	\draw[D1, postaction={decorate}] ({cos(30)*1.5}, {-1.5-1.5*sin(30)}) -- ++(30:1.5);
	\draw[D1, postaction={decorate}] ({2.5*cos(30)*1.5}, {-1.5-.75*sin(30)}) -- ++(150:.75);
	\draw[D1, postaction={decorate}] ({-cos(30)*.75}, {-1.5-.75*sin(30)}) -- ++(30:.75);
	\draw[D1, postaction={decorate}] ({cos(30)*1.5}, {sin(30)*1.5}) -- ++(0,.75);
	\draw[D1, postaction={decorate}] ({cos(30)*1.5}, {-1.5-2*sin(30)*1.5}) -- ++(0,.75);
\end{scope}
\begin{scope}[line cap=round,decoration={
	markings,
	mark=at position 0.5 with {\arrow{stealth}}}]
	\draw[red, postaction={decorate}] (.25,-.75) arc (180:-180:{sqrt(3)/2*1.5-.25});
\end{scope}

 \node[red] at (2,-0.8) {$s$};
 
\end{tikzpicture}}}.
     \label{eq:Bsp_Def}
\end{equation}

If any of the vertices around the plaquette violates the branching rules, then $B_p^s=0$ as $F$-symbols are non-zero only when branching rules are satisfied. In other words, the plaquette operator $B_p$ automatically includes a product of projectors $Q_v$ on the vertices $v$ around the plaquette, as discussed in~\cite{Levin_2005}. This leads to the aforementioned impossibility of creating isolated (pure) vertex excitations, since locally exciting a vertex will cause the surrounding plaquettes to be excited as well. Nonetheless, one can show that all $Q_v$ and $B_p$ operators commute and in addition that they are projectors, satisfying $Q_v^2=Q_v$ and $B_p^2=B_p$. Ground states are then simultaneous unit eigenvectors of all $B_p$ and $Q_v$ operators, such that quasiparticle excitations violate at least one of these projector constraints. These excitations can be labeled by simple objects of the Drinfeld center $\mathcal{Z}(\mathcal{C})$ of the input category (see e.g.~\cite{LanWen}). However, it will become apparent later in sections \ref{sec:TubeAlgebras} and \ref{sec:Degeneracies} that these excitations have subtypes that are not accounted for by $\dc$.

\subsection{Restricted Hilbert space: \texorpdfstring{$LW_0$}{Lg} model}\label{sec:LW0}
Instead of dealing with the full Hilbert space defined above, in this work we truncate the basis to the subspace where all vertices satisfy $Q_v=1$. This corresponds to the ``charge-free sector'' where there are no vertex excitations. Alternatively, it can also be thought of as an effective low-energy description of the general model in the limit $J_v \rightarrow \infty$. After setting the energy scale $J_p\equiv 1$, the Hamiltonian takes the simpler form
\begin{equation}
    H = -\sum_p B_p.
    \label{eq:Hamiltonian}
\end{equation}

In this construction, quasiparticle excitations located at a single plaquette form only a subset of the Drinfeld center $\mathcal{F}\subseteq \mathcal{Z}(\mathcal{C})$ that includes the vacuum and is called the \textit{fluxon} subset~\cite{Ritz-Zwilling2024_2, Ritz_Zwilling_2024}. However, the total quantum number emerging from a region containing two or more plaquettes can yield other anyons in $\dc$ that are composed from the fusion of multiple fluxons. This simpler version of the LW model is called $LW_0$ in this work, and has been studied previously, e.g. in~\cite{Ritz_Zwilling_2024,Ritz-Zwilling2024_2,Vidal2022}. It is nonetheless considered here for completeness and because it is the simplest model where the methodology developed in this work can be applied to.

In the rest of the work, the restricted Hilbert space is assumed for all models. The input graph can be described by its genus $g$ and the total number of plaquettes $N_p$, and the ground-state energy, which corresponds to specifying $B_p=1$ at all plaquettes, is given by $E_0=-N_p$.

\subsection{Extended string-net models \texorpdfstring{$LW_k$}{Lg} \label{sec:ExtendedModels}}
Extended string-net models, originally proposed in \cite{Hu_2018}, are slight modifications to the string-net model that allow for a unified treatment of all quasiparticle excitations described by $\mathcal{Z}(\mathcal{C})$, while staying in the restricted Hilbert space where all trivalent vertices respect the fusion rules of the input category. Typically~\cite{Hu_2018, MapMorita,Scipost_DiTc,Kawagoe2024}, extended  string-net models are constructed by associating to each vertex of the trivalent graph an open edge, called a tail, and attaching it to a neighbouring edge of the vertex. Each tail hosts an internal degree of freedom (called $q_p$) that takes values on simple objects of $\mathcal{C}$. In total, adding a tail adds two degrees of freedom: that on the tail and one on the attachment link that has been split in two. Variations on this extension scheme have also appeared in the literature \cite{Lin2024,Christian2023}.

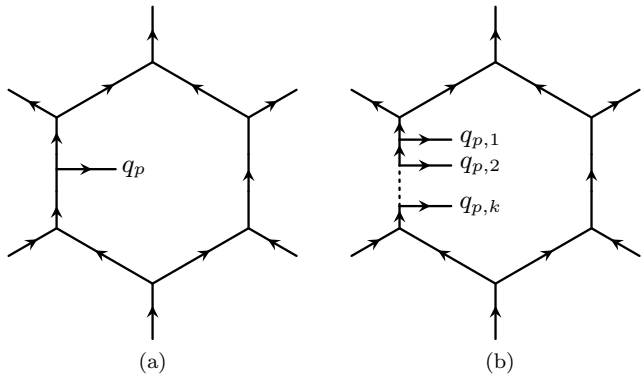
\begin{figure}
    \subfloat[\label{fig:HexagonTail}]{ \resizebox{!}{2.3cm}{\begin{tikzpicture}[thick,
		baseline={([yshift=-.5ex]current bounding box.center)},
		scale=.9
	]
\begin{scope}[line cap=round,decoration={
	markings,
	mark=at position 0.6 with {\arrow{stealth}}}]
	\draw[D1, postaction={decorate}] (0,-0.7) -- (.8,-0.7);
    \node[D1] at ($(.8,-0.7) + (0.25,0)$) {$q_p$};
\end{scope}
\begin{scope}[line cap=round,decoration={
	markings,
	mark=at position 0.6 with {\arrow{stealth}}}]
	
	\draw[D1, postaction={decorate}] (0,0) -- (30:1.5);
	\draw[D1, postaction={decorate}] (0,0) -- (150:.75);

	\draw[D1, postaction={decorate}] (0,-1.5) -- (0,-1);
	\draw[D1] (0,-1) -- (0,-.5);
	\draw[D1, postaction={decorate}] (0,-.5) -- (0,0);

	\draw[D1] ({cos(30)*1.5*2},-1.5) -- ({cos(30)*1.5*2},-1);
	\draw[D1, postaction={decorate}] ({cos(30)*1.5*2},-1) -- ({cos(30)*1.5*2},-.5);
	\draw[D1] ({cos(30)*1.5*2},-.5) -- ({cos(30)*1.5*2},0);
	
	\draw[D1, postaction={decorate}] ({cos(30)*3},0) -- ++(30:.75);
	\draw[D1, postaction={decorate}] ({cos(30)*3},0) -- ++(150:1.5);
	\draw[D1, postaction={decorate}] ({cos(30)*1.5}, {-1.5-1.5*sin(30)}) -- ++(150:1.5);
	\draw[D1, postaction={decorate}] ({cos(30)*1.5}, {-1.5-1.5*sin(30)}) -- ++(30:1.5);
	\draw[D1, postaction={decorate}] ({2.5*cos(30)*1.5}, {-1.5-.75*sin(30)}) -- ++(150:.75);
	\draw[D1, postaction={decorate}] ({-cos(30)*.75}, {-1.5-.75*sin(30)}) -- ++(30:.75);
	\draw[D1, postaction={decorate}] ({cos(30)*1.5}, {sin(30)*1.5}) -- ++(0,.75);
	\draw[D1, postaction={decorate}] ({cos(30)*1.5}, {-1.5-2*sin(30)*1.5}) -- ++(0,.75);
\end{scope}

\begin{scope}[line cap=round,decoration={
	markings,
	mark=at position 0.5 with {\arrow{stealth}}}]
\end{scope}
 
\end{tikzpicture}}}\hfill%
    \subfloat[\label{fig:HexagonKTail}]{\resizebox{!}{2.3cm}{\begin{tikzpicture}[thick,
		baseline={([yshift=-.5ex]current bounding box.center)},
		scale=.9
	]
\begin{scope}[line cap=round,decoration={
	markings,
	mark=at position 0.6 with {\arrow{stealth}}}]
    \draw[D1, postaction={decorate}] (0,-0.3) --++ (.7,0);
    \node[D1] at ($(.8,-0.3) + (0.3,0)$) {$q_{p,1}$};
    \draw[D1, postaction={decorate}] (0,-0.65) --++ (.7,0);
    \node[D1] at ($(.8,-0.65) + (0.3,0)$) {$q_{p,2}$};
    \draw[D1, postaction={decorate}] (0,-1.2) --++ (.7,0);
    \node[D1] at ($(.8,-1.2) + (0.3,0)$) {$q_{p,k}$};
\end{scope}
\begin{scope}[line cap=round,decoration={
	markings,
	mark=at position 0.6 with {\arrow{stealth}}}]
	
	\draw[D1, postaction={decorate}] (0,0) -- (30:1.5);
	\draw[D1, postaction={decorate}] (0,0) -- (150:.75);
	\draw[D1] ({cos(30)*1.5*2},-1.5) -- ({cos(30)*1.5*2},-1);
	\draw[D1, postaction={decorate}] ({cos(30)*1.5*2},-1) -- ({cos(30)*1.5*2},-.5);
	\draw[D1] ({cos(30)*1.5*2},-.5) -- ({cos(30)*1.5*2},0);
	
	\draw[D1, postaction={decorate}] ({cos(30)*3},0) -- ++(30:.75);
	\draw[D1, postaction={decorate}] ({cos(30)*3},0) -- ++(150:1.5);
	\draw[D1, postaction={decorate}] ({cos(30)*1.5}, {-1.5-1.5*sin(30)}) -- ++(150:1.5);
	\draw[D1, postaction={decorate}] ({cos(30)*1.5}, {-1.5-1.5*sin(30)}) -- ++(30:1.5);
	\draw[D1, postaction={decorate}] ({2.5*cos(30)*1.5}, {-1.5-.75*sin(30)}) -- ++(150:.75);
	\draw[D1, postaction={decorate}] ({-cos(30)*.75}, {-1.5-.75*sin(30)}) -- ++(30:.75);
	\draw[D1, postaction={decorate}] ({cos(30)*1.5}, {sin(30)*1.5}) -- ++(0,.75);
	\draw[D1, postaction={decorate}] ({cos(30)*1.5}, {-1.5-2*sin(30)*1.5}) -- ++(0,.75);
\end{scope}

\begin{scope}[line cap=round,decoration={
	markings,
	mark=at position 0.8 with {\arrow{stealth}}}]

    \draw[D1, postaction=decorate] (0,-1.5) -- (0,-1.2);
    \draw[D1, dotted] (0,-1.2) -- (0,-.65);
    \draw[D1, postaction=decorate] (0,-.65) -- (0,-0.3);
    \draw[D1, postaction=decorate] (0,-.3) -- (0,0);
\end{scope}
 
\end{tikzpicture}}}
\caption{Graphical interpretation of appending tails to an hexagonal plaquette: (a) Plaquette with a single tail of the $LW_1$ model; (b) Plaquette with $k$ tails of generalized models $LW_k$.}
\label{fig:Hexagons_with_Tails}
\end{figure}

In the present work we slightly shift the perspective as we choose to associate tails to the plaquettes, instead of the vertices, of the input graph. Thus, for each plaquette, the graph is extended by attaching one tail to it, meaning that the tail penetrates to the middle of the plaquette. The choice of which edge of the plaquette the tail is attached to is not important. An example of this extension to a hexagonal plaquette is shown in Fig.~\ref{fig:HexagonTail}. The tail corresponds to a dangling string and, when it is not in the vacuum state, it is able to simulate the violation of a vertex constraint.

The Hamiltonian for this model is the same as \eqref{eq:Hamiltonian}, but now the $B_p$ operators are slightly modified to accommodate for the added tails. Their internal structure is detailed in \cite{Hu_2018}. In short, if $q_p=1$, they act in the same way as shown in Eq. \eqref{eq:Bsp_Def} and if $q_p\neq 1$, they yield $B_p=0$. 

It was shown in different contexts \cite{Hu_2018,Christian2023,Kawagoe2024} that the addition of a single tail to a plaquette is enough to generate all the excitations of $\mathcal{Z}(\mathcal{C})$ at the single plaquette level, contrary to what is known for the $LW_0$ model. An important consequence is that all anyons will now appear as plaquette excitations. While some plaquette excitations have an un-excited tail (i.e. the tail is in the vacuum state) and correspond to the fluxons, other plaquette excitations have an excited tail, which is typically interpreted as a charge quantum number (i.e., as if a vertex was excited).

One can also devise models where each plaquette is associated to more than one tail. As such, we further define a family of models which we will call $LW_k$, where each plaquette has $k \in \mathbb{N}$ tails attached to it, see Fig.~\ref{fig:HexagonKTail}. In particular, on a torus (genus $g=1$), the $LW_2$ model is equivalent to the extended string-net model proposed by Hu, Geer and Wu~\cite{Hu_2018} in which each tail corresponds to a vertex (indeed, a trivalent graph on a torus has $N_v = 2 N_p$). There is also the possibility of constructing models where plaquettes have distinct numbers of tails. Although the ideas developed in this work can be applied to these cases, they will not be mentioned further because they do not bring additional features.

\section{\label{sec:TubeAlgebras} Tube algebras}
Starting from an input category $\mathcal{C}$ that is a UFC, the tube algebra can be used to construct the Drinfeld center $\mathcal{Z(C)}$, which is a UMTC (for review, see e.g.~\cite{Simon2023}).  A UMTC is a UFC with additional structure, namely braiding and an extra condition that we detail below.  The Drinfeld center is made of simple objects $A$ (called anyons), that obey fusion rules such as $A\times B = \sum_C N_{A,B}^C \, C$ and have quantum dimensions $d_A$.  That the fusion category is braided means that it also has $S$- and $T$- matrices that encode the mutual-statistics $S_{A,B}$ and self-statistics $T_{A,A}$ of the anyons.  In addition, the $S$-matrix is modular, i.e. it is unitary, $S S^\dagger = S^\dagger S= \mathds{1}$, which means that the braiding of anyons is well-defined and not degenerate. The total quantum dimension of the Drinfeld center is \be
\mathcal{D}=\sqrt{\sum_{A\in \mathcal{Z(C)}} d_A^2}=\mathcal{D_C}^2,
\ee
in terms of the total quantum dimension of the input category $\mathcal{C}$.

Excitations of string-net models are labeled by the simple objects of the Drinfeld center $A \in \mathcal{Z}(\mathcal{C})$, which are in one-to-one correspondence with the simple (irreducible) modules of the \emph{tube algebra} \cite{LanWen,Hu_2018}. 

In this section, we will review the structure of a family of tube algebras, classified by the number $k$ of open-ended strings (or tails) which will act on the different extended $LW_k$ models. We will also introduce the Morita equivalence between them, highlighting the importance of the tube algebras with $k=0$ and $k=1$. 

\subsection{\label{subsec:Family_of_TA} \texorpdfstring{$k$-}{Lg}tails tube algebras \texorpdfstring{$\mathcal{TA}_k$}{Lg}}

\begin{figure}[ht!]
    \subfloat[\label{fig:Tube1}]{\resizebox{!}{1.5cm}{\begin{tikzpicture}[thick,
		baseline={([yshift=-.5ex]current bounding box.center)},
		scale=.9
	]
\begin{scope}[line cap=round,decoration={
	markings,
	mark=at position 0.9 with {\arrow{stealth}}},
  line width= 0.5pt]

	\draw[D1, postaction={decorate}] (.73,-1.2) --++ (.5,0);
\end{scope}

\begin{scope}[line cap=round,decoration={
	markings,
	mark=at position 0.5 with {\arrow{stealth}}},
  line width= 0.5pt]
   \draw[D1, postaction={decorate}] (.35,-0.2) --++ (.45,0);
	\draw[D1, postaction={decorate}] (.6,-.75) arc (180:-180:{sqrt(3)/2*1.5-.45});
\end{scope}

\begin{scope}[line cap=round,decoration={
	markings,
	mark=at position 0.01 with {\arrow{stealth}}},
  line width= 0.5pt]
	\draw[D1, postaction={decorate}] (.6,-.75) arc (180:-180:{sqrt(3)/2*1.5-.45});
\end{scope}

 \node[D1] at (1.4,-1.2) {$s$};
 \node[D1] at (.4,-0.05) {$r$};

 \node[D1] at (2,-0.6) {$i$};
 \node[D1] at (0.4,-0.8) {$j$};
 
\end{tikzpicture}}}\hfill%
    \subfloat[\label{fig:Tube2}]{\resizebox{!}{1.5cm}{\begin{tikzpicture}[thick,
		baseline={([yshift=-.5ex]current bounding box.center)},
		scale=1.,
		D1
	]
\begin{scope}[line cap=round,decoration={
	markings,
	mark=at position 0.4 with {\arrow{stealth}}}]

	\draw[D1, postaction={decorate}] (.55-0.5,.2) -- (.55,.2);
	\draw[D1, postaction={decorate}] (.28-0.5,-1) -- (.28,-1);

\end{scope}

\begin{scope}[line cap=round,decoration={
	markings,
	mark=at position 0.9 with {\arrow{stealth}}}]
	\draw[D1, postaction={decorate}] (.28,-.4) --($(.28,-.4)+(.5,0)$);
	\draw[D1, postaction={decorate}] (.5,-1.6) -- ($(.5,-1.6)+(.5,0)$);

\end{scope}

\begin{scope}[line cap=round,decoration={
	markings,
	mark=at position 0.5 with {\arrow{stealth}}}]
	\draw[D1, postaction={decorate}] (.25,-.75) arc (180:-180:1.5);
\end{scope}

\begin{scope}[line cap=round,decoration={
	markings,
	mark=at position 0.08 with {\arrow{stealth}}}]
	\draw[D1, postaction={decorate}] (.25,-.75) arc (180:-180:1.5);
\end{scope}

\begin{scope}[line cap=round,decoration={
	markings,
	mark=at position 0.02 with {\arrow{stealth}}}]
	\draw[D1, postaction={decorate}] (.25,-.75) arc (180:-180:1.5);
\end{scope}

\begin{scope}[line cap=round,decoration={
	markings,
	mark=at position 0.95 with {\arrow{stealth}}}]
	\draw[D1, postaction={decorate}] (.25,-.75) arc (180:-180:1.5);
\end{scope}

\node[black] at (.05-0.14,.2+0.15) {$a$};
\node[black] at (.28-0.5-0.14,-1+0.15) {$b$};
\node[black] at ($(.28,-.4)+(.5,0) + (0.15,0)$) {$c$};
\node[black] at ($(.5,-1.6)+(.5,0) + (0.15,0)$) {$d$};

\node[black] at (3.5,-0.5) {$e$};
\node[black] at (0.2,-0.05) {$f$};
\node[black] at (0.08,-0.75) {$g$};
\node[black] at (0.18,-1.35) {$h$};

\end{tikzpicture}}}
\caption{(a) Example of a tube $\mathcal{T}^{i;j}_{r;s}\in \mathcal{TA}_1$ $=Q$-algebra. (b) Example of a tube of the 2-tails algebra, labelled by 8 internal strings $\mathcal{T}_{ab;cd}^{e;fgh}\in \mathcal{TA}_2$ $=\phi$-algebra.
}
\label{fig:ExampleTubes}
\end{figure}
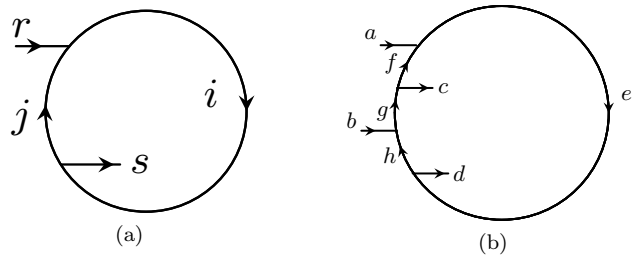

The elements of the tube algebra are called tubes $\mathcal{T}_{\{i\}}$,  and are built from a set of strings $\{i\}\in \mathcal{C}$ drawn on an annulus or cylinder. Some of these strings have no open ends and together wind around the periodic direction of the tube (see, e.g., the strings labeled $i$ and $j$ in Fig.~\ref{fig:Tube1}), while others have an open end at either the top (i.e., the inner part of the annulus, see e.g. the string $s$ in Fig.~\ref{fig:Tube1}) or bottom (i.e., the outer part of the annulus, see e.g. the string $r$ in Fig.~\ref{fig:Tube1}) of the tube. These open-ended strings will be called ``tails" in the following. All fusion vertices inside the tube must be stable. For example, if there is one tail on each side of the tube, then all the tubes can be indexed by four labels $\mathcal{T}^{i;j}_{r;s}, \{i,r,s,j\}\in \mathcal{C}$ forming a diagram as shown in Fig.~\ref{fig:Tube1}. In this example, there are two vertices at which the fusion rules must be satisfied. 

The tubes can be composed by stacking them together, gluing the tails maintaining their order and using the $F$-symbols to rearrange the resulting structure into a combination of tubes:
\begin{equation}
    \mathcal{T}_{\{i\}}\mathcal{T}_{\{j\}} = \sum_{\{k\}} f_{\{i\}\{j\}}^{\{k\}} \mathcal{T}_{\{k\}},
\end{equation}
where $\{i\}$ is a set of input strings that characterizes each tube and $f_{\{i\}\{j\}}^{\{k\}}$ are the structure constants.  As this composition requires matching the tubes' tails in a specific order, the collection of tubes with the same number of tails form a closed algebra under the above composition. Thus, we can construct different tube algebras $\mathcal{TA}_k$ indexed by the number $k\in \mathbb{N}$ of tails on each side of the tubes. We call $\mathcal{TA}_0$ the $O$-algebra, $\mathcal{TA}_1$ the $Q$-algebra and $\mathcal{TA}_2$ the $\phi$-algebra, because of the shape of the tubes in each algebra. For a complete introduction to the $Q$-algebra see e.g.~\cite{LanWen,Simon2023,Ritz_Zwilling_2024}.

The number of tubes in each $\mathcal{TA}_k$ is equal to the dimension of the algebra. Denoting the set of tails on the outside/inside of the tube by $\{o_1,\dots,o_k\}$/$\{i_1,\dots,i_k\}$ and using the fact that all internal vertices of the tubes are stable, we obtain:
\begin{equation}
    \dim (\mathcal{TA}_k) = \sum_{\{o_1,\dots,o_k\}}\sum_{\{i_1,\dots,i_k\}} \text{Tr}\left(\prod_{j=1}^k N_{o_j}N_{i_j}\right),
    \label{eq:DimTAk_1}
\end{equation}
where $N _{o_j},N_{i_j}$ are fusion matrices of $\mathcal{C}$.

Slightly different perspectives on the role of the tube algebra in the context of string-net models have appeared in the literature. For instance, the action of the tubes can be viewed as ``gluing" a tube of ground-state around a region containing a quasiparticle excitation~\cite{Simon2023,LanWen,Bullivant2019}. But the tubes can also be thought of as local operators acting directly in the plaquettes~\cite{Hu_2018,MapMorita}, which is the perspective we will adopt. The action of a tube $\mathcal{T}_{\{i\}}(p)$ on a plaquette $p$ can be viewed as inserting the corresponding diagram inside a plaquette (see Fig.~\ref{fig:ExampleTubes}), connecting the tails of the plaquette with the tails of the tube and fusing the resulting diagram to the boundary of the plaquette.

Since one needs to connect the tails of the plaquette with the open strings of the tube, it is clear that the set of tubes that can act on the plaquettes of the $LW_k$ model is the same set that composes $\mathcal{TA}_k$. In turn, this means that the excitations of the $LW_k$ model are described by the simple modules of $\mathcal{TA}_k$. For example, the action of a tube $\mathcal{T}^{i;j}_{r;s} \in Q$-algebra on a plaquette $p$ of the $LW_1$ model can be graphically depicted as:
\begin{equation}
    \mathcal{T}^{i;j}_{r;s}(p)\ket{ \resizebox{!}{1.5cm}{\begin{tikzpicture}[thick,
		baseline={([yshift=-.5ex]current bounding box.center)},
		scale=.9
	]
\begin{scope}[line cap=round,decoration={
	markings,
	mark=at position 0.6 with {\arrow{stealth}}}]
	\draw[D1, postaction={decorate}] (0,-0.7) -- (.8,-0.7);
    \node[D1] at ($(.8,-0.7) + (0.25,0)$) {$q_p$};
\end{scope}
\begin{scope}[line cap=round,decoration={
	markings,
	mark=at position 0.6 with {\arrow{stealth}}}]
	
	\draw[D1, postaction={decorate}] (0,0) -- (30:1.5);
	\draw[D1, postaction={decorate}] (0,0) -- (150:.75);

	\draw[D1, postaction={decorate}] (0,-1.5) -- (0,-1);
	\draw[D1] (0,-1) -- (0,-.5);
	\draw[D1, postaction={decorate}] (0,-.5) -- (0,0);

	\draw[D1] ({cos(30)*1.5*2},-1.5) -- ({cos(30)*1.5*2},-1);
	\draw[D1, postaction={decorate}] ({cos(30)*1.5*2},-1) -- ({cos(30)*1.5*2},-.5);
	\draw[D1] ({cos(30)*1.5*2},-.5) -- ({cos(30)*1.5*2},0);
	
	\draw[D1, postaction={decorate}] ({cos(30)*3},0) -- ++(30:.75);
	\draw[D1, postaction={decorate}] ({cos(30)*3},0) -- ++(150:1.5);
	\draw[D1, postaction={decorate}] ({cos(30)*1.5}, {-1.5-1.5*sin(30)}) -- ++(150:1.5);
	\draw[D1, postaction={decorate}] ({cos(30)*1.5}, {-1.5-1.5*sin(30)}) -- ++(30:1.5);
	\draw[D1, postaction={decorate}] ({2.5*cos(30)*1.5}, {-1.5-.75*sin(30)}) -- ++(150:.75);
	\draw[D1, postaction={decorate}] ({-cos(30)*.75}, {-1.5-.75*sin(30)}) -- ++(30:.75);
	\draw[D1, postaction={decorate}] ({cos(30)*1.5}, {sin(30)*1.5}) -- ++(0,.75);
	\draw[D1, postaction={decorate}] ({cos(30)*1.5}, {-1.5-2*sin(30)*1.5}) -- ++(0,.75);
\end{scope}

\begin{scope}[line cap=round,decoration={
	markings,
	mark=at position 0.5 with {\arrow{stealth}}}]
\end{scope}
 
\end{tikzpicture}}} = \delta_{r,q_p} \ket{ \resizebox{!}{1.5cm}{\begin{tikzpicture}[thick,
		baseline={([yshift=-.5ex]current bounding box.center)},
		scale=.9
	]
\begin{scope}[line cap=round,decoration={
	markings,
	mark=at position 0.8 with {\arrow{stealth}}}]
	\draw[D1] (0,-0.7) -- (.28,-0.7);
	\draw[D1, postaction={decorate}] (.73,-1.2) --++ (.5,0);
\end{scope}
\begin{scope}[line cap=round,decoration={
	markings,
	mark=at position 0.6 with {\arrow{stealth}}}]
	
	\draw[D1, postaction={decorate}] (0,0) -- (30:1.5);
	\draw[D1, postaction={decorate}] (0,0) -- (150:.75);

	\draw[D1, postaction={decorate}] (0,-1.5) -- (0,-1);
	\draw[D1] (0,-1) -- (0,-.5);
	\draw[D1, postaction={decorate}] (0,-.5) -- (0,0);

	\draw[D1] ({cos(30)*1.5*2},-1.5) -- ({cos(30)*1.5*2},-1);
	\draw[D1, postaction={decorate}] ({cos(30)*1.5*2},-1) -- ({cos(30)*1.5*2},-.5);
	\draw[D1] ({cos(30)*1.5*2},-.5) -- ({cos(30)*1.5*2},0);
	
	\draw[D1, postaction={decorate}] ({cos(30)*3},0) -- ++(30:.75);
	\draw[D1, postaction={decorate}] ({cos(30)*3},0) -- ++(150:1.5);
	\draw[D1, postaction={decorate}] ({cos(30)*1.5}, {-1.5-1.5*sin(30)}) -- ++(150:1.5);
	\draw[D1, postaction={decorate}] ({cos(30)*1.5}, {-1.5-1.5*sin(30)}) -- ++(30:1.5);
	\draw[D1, postaction={decorate}] ({2.5*cos(30)*1.5}, {-1.5-.75*sin(30)}) -- ++(150:.75);
	\draw[D1, postaction={decorate}] ({-cos(30)*.75}, {-1.5-.75*sin(30)}) -- ++(30:.75);
	\draw[D1, postaction={decorate}] ({cos(30)*1.5}, {sin(30)*1.5}) -- ++(0,.75);
	\draw[D1, postaction={decorate}] ({cos(30)*1.5}, {-1.5-2*sin(30)*1.5}) -- ++(0,.75);
\end{scope}
\begin{scope}[line cap=round,decoration={
	markings,
	mark=at position 0.5 with {\arrow{stealth}}}]
	\draw[D1, postaction={decorate}] (.6,-.75) arc (180:-180:{sqrt(3)/2*1.5-.45});
    \draw[D1, postaction={decorate}] (.28,-0.7) --++ (.3,0);
\end{scope}

\begin{scope}[line cap=round,decoration={
	markings,
	mark=at position 0.97 with {\arrow{stealth}}}]
	\draw[D1, postaction={decorate}] (.6,-.75) arc (180:-180:{sqrt(3)/2*1.5-.45});
\end{scope}

 \node[D1] at (1.4,-1.2) {$s$};
 \node[D1] at (.4,-0.5) {$q_p$};

 \node[D1] at (2,-0.6) {$i$};
 \node[D1] at (0.45,-1) {$j$};
 
\end{tikzpicture}}}.
    \label{eq:TubeAction}
\end{equation}

With this perspective, the operators $B_p^s$ defined in Eq.~\eqref{eq:Bsp_Def} are equivalent to the action of tubes with trivial tails, corresponding to $\mathcal{T}_{1;1}^{s;s}(p)$ for the $Q$-algebra.

\subsection{\label{subsec:Idempotents_and_Multiplicities} Algebra idempotents and internal multiplicities}

By the Artin-Wedderburn theorem (cf. Appendix \ref{appendix:TAdecomp}), each of these tube algebras can be decomposed into a direct sum of simple matrix algebras,  each of dimension $n_A$. Concretely, one is interested in special objects that project onto each irreducible block $A$, encoded in the simple idempotents $p_A^{\alpha\alpha}$ and nilpotents $p_A^{\alpha\beta}$. An idempotent satisfies $(p_A^{\alpha\alpha})^2=p_A^{\alpha\alpha}$ and is simple if it cannot be written as a sum of two other idempotents.  A nilpotent is such that $(p_A^{\alpha\beta})^2=0$. Symbolically, they are defined as:
\begin{equation}
    p^{\alpha\beta}_A=\sum_{\{i\}} \left(M^{\alpha\beta}_{A,\,\{i\}}\right)^{-1} \mathcal{T}_{\{i\}} \quad \quad \quad 
    p^{\alpha\beta}_A p^{\mu\nu}_B = \delta_{A,B}\delta_{\beta,\mu} p^{\alpha \nu}_A,
    \label{eq:IdempotentDef}
\end{equation}
where we have defined the tube algebra's tensors $M^{\alpha\beta}_{A,\{i\}}$ that relate the tubes to the simple idempotents and nilpotents. The tube algebra's tensors $M_A$ are proportional to the half-braiding tensors $\Omega_A$~\cite{LanWen}. Each irreducible block $A$ corresponds to an anyon of the Drinfeld center, while the simple idempotents carry are defined within each block and carry an extra label $\alpha$ associated to the different ``subtypes" of the corresponding anyon. 
A simple idempotent is such that
\begin{equation}
    n_{A,\alpha}=\text{Tr}(p_A^{\alpha \alpha})= 1 \text{ or }0.
    \label{eq:InternalMult_def}
\end{equation}
When $n_{A,\alpha}$ vanishes, it means that $p_A^{\alpha \alpha}$ does not exist.

For the $Q$-algebra of a UFC $\mathcal{C}$ with commutative fusion rules, $\alpha \equiv s\in \mathcal{C}$ takes values on the input strings (c.f. appendix \ref{appendix:SpecialSubset}). One can then interpret each internal multiplicity $n_{A,s}=0$ or $1$ as the number of type $s$ strings that $A$ decomposes into. 

Still for the $Q$-algebra, if $\mathcal{C}$ is a non-commutative UFC then the simple idempotents are indexed by $s\in \mathcal{C}$ and an extra label $a\in \mathbb{N}^*$ such that $\alpha \equiv (s,a)$~\cite{Ritz_Zwilling_2024}. In this case, one can build idempotents:
\begin{equation}
    P_A^{ss} = \sum_a p_A^{(s,a)(s,a)},
    \label{eq:NonCommutative_Idempotents}
\end{equation}
which are not simple, but whose dimensions
\begin{equation}
 n_{A,s}=\text{Tr}(P_A^{s s})=\sum_a n_{A,(s,a)} \in \mathbb{N}
    \label{eq:NonCommutative_InternalMult}
\end{equation}
again count the number of type $s$ strings in $A$. Note that $n_{A,s}> 1$ is now possible. Generally, for $\mathcal{TA}_k$, $k \geq 2$ the labels $\alpha$ have more structure.

To obtain irreducible representations of the tube algebra, the simple idempotents have to be combined into simple central idempotents $P_A$. These are the projectors onto each topological quasiparticle $A$, which commute with all the tubes and satisfy the following relations:
\begin{equation}
    P_A = \sum_\alpha p_A^{\alpha\alpha} \quad \quad \left[ P_A, \mathcal{T}_{\{i\}}\right]=0 \quad \quad \text{Tr}\left( P_A \right) = n_A.
    \label{eq:CentralIdemp_def}
\end{equation}

These simple central idempotents are in a one-to-one correspondence with the simple objects $A$ of the Drinfeld center $\mathcal{Z}(\mathcal{C})$. In particular, for the $Q$-algebra, when $A=\mathbf{1}$ is the vacuum we have $P_{\mathbf{1}}=B_p$~\footnote{To avoid confusion, $1\in \mathcal{C}$  is used for the vacuum of the input category and the bold font $\mathbf{1}\in \dc$ for the vacuum of its Drinfeld center.}, which is equal to the ``Kirby strand" acting around a plaquette~\cite{Simon2023}.

The right-most equality of \eqref{eq:CentralIdemp_def} states that the dimension of each central idempotent of the tube algebra is characterized by a non-negative integer $n_A$. From \eqref{eq:InternalMult_def}, one obtains that
\begin{equation}
    n_A = \sum_\alpha n_{A,\alpha}.
    \label{eq:Multiplicities}
\end{equation}

These dimensions, sometimes referred to as internal multiplicities, will play a crucial role in the counting of level degeneracies in string-net models because they are necessary to fully describe the emergent anyon excitations.  They encode the number of different anyon subtypes, which are associated to different microscopic configurations of the string-net graph.

\subsection{\label{subsec:MoritaEquivalence}Morita equivalence and importance of the \texorpdfstring{$Q$}{Lg}-algebra}
In order to study the anyon excitations, one is interested in the tube algebra's simple modules (i.e. the irreducible representations) rather than in the algebra itself. Indeed, the knowledge of the tube algebra's tensors $M_{A,\{i\}}^{\alpha \beta}$ (closely related to the half-braiding tensors~\cite{LanWen}) is sufficient to obtain the modular data of the output topological order $\dc$. 
If two distinct algebras $\mathcal{A}$ and $\mathcal{A}^\prime$ have equivalent simple modules, then the excitations can be labelled using either of them and the two algebras are said to be \textit{Morita equivalent}~\cite{Morita1958DualityFM,Kong2012}~\footnote{As a warning to the reader, we note that the notion of Morita equivalence can also be extended to UFCs: two distinct fusion categories $\mathcal{C}$ and $\mathcal{C}^\prime$ are said to be Morita equivalent if they have the same Drinfeld center $\mathcal{Z}(\mathcal{C})\cong \mathcal{Z}(\mathcal{C}^\prime)$~\cite{Mueger2003}.}.

In the present work, we will be especially interested in the Morita equivalence between the different tube algebras $\mathcal{TA}_k$, $k\geq1$~\cite{LanWen} built from the same input category (UFC) $\mathcal{C}$. This implies that, in order to obtain the quasiparticle excitations and the modular data that characterizes the emergent topological order of the extended string-net models, it is enough to decompose the simplest algebra $\mathcal{TA}_1$ (the $Q$-algebra). By contrast, the simple modules of  $\mathcal{TA}_0$, whose tubes correspond to closed loops of input strings $s\in \mathcal{C}$, only form a subset $\mathcal{F}\subseteq \mathcal{Z}(\mathcal{C})$ called the \textit{fluxon} subset~\cite{Ritz_Zwilling_2024}. Therefore, $\mathcal{TA}_0$ is not in the same Morita equivalence class as the other tube algebras.

Despite the equivalence of simple modules of this family of tube algebras, the structure of the central idempotents changes according to the number $k$ of tails. In particular, it is not hard to see that the dimensions of the central idempotents must increase with $k$. On the one hand, the dimension of $\mathcal{TA}_k$ is equal to the number of tubes, as given in equation \eqref{eq:DimTAk_1}. On the other hand, since each simple matrix algebra of the Artin-Wedderburn decomposition has dimension $n_A$, it follows that:
\begin{equation}
    \dim (\mathcal{TA}_k) = \sum_A n_A^2.
\label{eq:DimTAk_2}
\end{equation}

Equations (\ref{eq:DimTAk_1},~\ref{eq:DimTAk_2}) imply that the internal multiplicities $n_A$ must increase with the parameter $k$, so we will denote $n_A^{(k)}$ the multiplicities arising from $\mathcal{TA}_k$ and reserve the notation $n^Q_A$ for $k=1$. In particular, for $k=1$, Eq.~\eqref{eq:Multiplicities} reads:
\begin{equation}
    n_A^{Q}= \sum_{s\in \mathcal{C}} n^Q_{A,s}.
\end{equation}

To obtain the integers $n_A^{(k)}$ when $k\geq2$, there is a shortcut that does not involve the full idempotent decomposition of the tube algebra. Indeed, these multiplicities for $k >1$ can be obtained directly from $n_{A,s}^Q$ using the formula:
\begin{equation}
    n_A^{(k)} = \sum_{s\in \mathcal{C}} n^Q_{A,s} \left[ \left( \sum_{t\in \mathcal{C}} N_t \right)^k\right]_{1,s},
    \label{eq:Nk_from_N1}
\end{equation}
where the last subscript refers to row 1 and column $s$ of the matrix inside the brackets. Although this expression does not give the tube algebra's tensors, it provides the necessary information to compute the partition function of the extended string-net models, as will become clear in the following sections. 
As was defined in~\cite{Ritz_Zwilling_2024}, the fluxons $\mathcal{F}$ can also be directly identified from the multiplicities of the $Q$-algebra, since they are the simple objects $F \in \mathcal{Z}(\mathcal{C})$ that satisfy $n^Q_{F,1} \geq 1$.

Because the $Q$-algebra plays a central role, in the following sections we will focus on the $LW_1$ model, from which the results can be generalized to $LW_k$.

\section{\label{sec:Degeneracies} Spectral degeneracies and partition function}
In this section, the degeneracies of  excited levels in extended string-net models are obtained. The local tube algebra of plaquette operators is used to identify the different microscopic configurations that violate the $B_p=1$ constraints. With an analytic expression for the spectral degeneracies, we get access to the partition function and to the finite temperature properties. We deal explicitly with the $LW_1$ model throughout, and then explain how it generalizes to the family of extended string-net models $LW_k$ presented in the previous sections.

\subsection{\label{subsec:Spectral_Degeneracies} Spectral degeneracies}
There are two kinds of degeneracies arising in string-net models, which we call topological and non-topological. Topological degeneracies, normally associated to the ground-state degeneracy of topological orders, are protected against any type of small, local perturbations. However, lattice models of topological orders display additional degeneracies, which do not have a topological nature and can be lifted by local perturbations.

\subsubsection{Topological degeneracies}
Topological degeneracies were first worked out in~\cite{Moore_1989} and depend only on the emergent topological order, characterized by an UMTC $\mathcal{U}$, and on the topology of the orientable and closed 2D manifold where the system is embedded in, labeled by the genus $g$. The problem of computing the Hilbert space dimension associated to a TQFT with some excitations $A_n\in \mathcal{U}$ can be mapped onto a problem of counting the possible stable fusion diagrams embedded on the corresponding manifold, with open strings labeled with the excitations $A_n$ \cite{Ritz-Zwilling2024_2,Moore_1989}.
\begin{figure}
    \centering
    \resizebox{!}{2cm}{\begin{tikzpicture}[thick,decoration={
		markings,
		mark=at position 0.6 with {\arrow{stealth}}},
		baseline={([yshift=-.5ex]current bounding box.center)},
		line cap=round
	] 
	
\begin{scope}[black]

    \draw[D1, postaction={decorate}] (135:2) -- (135:1.5);
    \draw[D1, postaction={decorate}] (135:1.5) -- (135:0.5);
    \draw[D1, postaction={decorate}] (135:0.5) -- (-45:0.5);
    \draw[D1, postaction={decorate}] (-45:0.5) -- (-45:1.5);
    \draw[D1] (-45:1.5) -- (-45:2);
    
    \draw[D1, postaction={decorate}] ($(-45:2) + (135:3.5) +(45:0.5)$) -- ($(-45:2) + (135:3.5)$);
    \draw[D1, postaction={decorate}] ($(-45:2) + (135:2.5) +(45:1.5)$) -- ($(-45:2) + (135:2.5)$);
    
    \node[D1] at  ($(135:2) + (-0.15,0.15)$) {$A_1$};
    \node[D1] at  ($(-45:2) + (135:3.5) +(45:0.5)+ (0.15,0.15)$) {$A_2$};
    \node[D1] at  ($(-45:2) + (135:3.5) +(45:0.5)+ (0.8,0.15)$) {$\dots$};
    \node[D1] at  ($(-45:2) + (135:2.5) +(45:1.5)+ (0.15,0.15)$) {$A_m$};
    
\end{scope}

\begin{scope}[black] 
    \draw[D1, postaction={decorate}] ($(-45:2) + (135:1.5) + (45:2.45)$) --++($(45:-2.45)$);
    \draw[D1, postaction={decorate}] ($(-45:2) + (135:.5) + (45:3.45)$) --++ ($(45:-3.45)$);
    \draw[D1, postaction={decorate}] ($(-45:2) + (45:3.95)$) --++ ($(45:-3.95)$);

    \draw[D1]  ($(-45:2) + (135:1.5) + (45:2.7)$) circle (0.24);
    \draw[D1]  ($(-45:2) + (135:.5) + (45:3.7)$)  circle (0.24);
    \node[D1] at  ($(-45:2) + (45:3.5) +(135:0.5) + (- 0.6,0.18)$) {\dots};
    \draw[D1]  ($(-45:2) + (45:4.2)$) circle (0.24);

    \draw[D1] [decorate,
    decoration = {brace}] ($(-45:2) + (135:1.5) + (45:2.45) + (-0.2,0.6)$) --   ($(-45:2) + (45:3.95) + (0.6,0.6)$);
    \node[align=left] at (3.4,2.3) {$g$ holes};
\end{scope}

\end{tikzpicture}}
    \caption{Fusion tree used to compute the Hilbert space dimension. On the left-most branches, the simple objects $A_n\in \mathcal{Z}(\mathcal{C})$ denote the excitations, while the others account for the topology of the manifold. All internal trivalent vertices are stable and the unlabeled edges are summed over in the computation.}
    \label{fig:MSB_Fusion}
\end{figure}
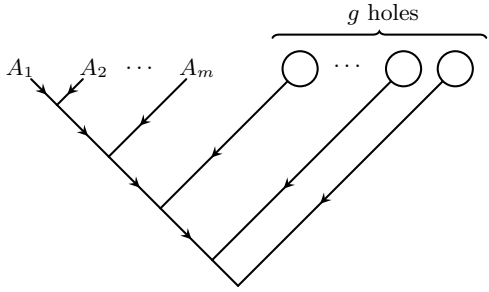

Denote with $\dim_{\mathcal{U}}\left(g,\{A_n\}_{n\leq m}\right)$ the Hilbert space dimension corresponding to the UMTC $\mathcal{U}$ embedded on a manifold of genus $g$ containing $m$ excitations labeled with $A_n$. Using $F$-moves, any planar fusion diagram can be restructured into the form of Fig.~\ref{fig:MSB_Fusion} without changing the corresponding Hilbert space dimension. The dimension of each stable internal vertex is given by $N_{AB}^{C}$ (i.e. the fusion matrix elements of the UMTC $\mathcal{U}$), and the total dimension is given by summing over all internal labels. Using the Verlinde formula~\cite{Verlinde_1988}:
\begin{equation}
    N_{AB}^{C} = \sum_{D\in \mathcal{U}} \frac{S_{A,D}S_{B,D}S_{\bar{C},D}}{S_{\mathbf{1},D}},
\end{equation}
and the fact that the $S$-matrix of a UMTC is both symmetric and unitary, one obtains the Moore-Seiberg-Banks formula~\cite{Moore_1989}:
\begin{equation}
    \dim_{\mathcal{U}}\left(g,\{A_n\}_{n\leq m}\right) = \sum_{C\in\mathcal{U}} S_{\textbf{1},C}^{2-2g-m} \prod_{n=1}^m S_{A_n,C}.
    \label{eq:MSB}
\end{equation}

\subsubsection{Non-topological degeneracies}
In string-net models, the emergent topological order $\mathcal{U}$ corresponds to the Drinfeld center $\mathcal{Z}(\mathcal{C})$ of an input UFC $\mathcal{C}$. However, making this substitution in the previous formulas is not enough, since Eq.~\eqref{eq:MSB} does not account for all the degrees of freedom of the lattice model. Indeed, there can be several microscopical configurations associated to each fusion tree of Fig.~\ref{fig:MSB_Fusion} because each anyon excitation $A_n$ can appear in distinct forms or subtypes.

To see why this is the case, first notice that the anyon excitations $A_n$ on the $LW_1$ model are hosted at the plaquettes of the input graph. Let $\ket{\psi}$ denote a state, corresponding to a particular microscopical configuration of the input graph, with an anyon excitation $B\in \mathcal{Z}(\mathcal{C})$ at a plaquette $p$. Because we can identify the tubes of the $Q$-algebra with local operators $\mathcal{T}^{i;j}_{r;s}(p)$ that act locally on each plaquette $p$, we may identify projection operators
\begin{equation}
     \Pi_{A}^{(s,a) (s,a)}(p) = \sum_{ij \in \mathcal{C}} \sum_{a=1}^{n_{A,s}} \left(M_{A,\, issj}^{(s,a) (s,a)} \right)^{-1} \mathcal{T}^{i;j}_{s;s}(p),
     \label{eq:QAlgebra_Modules}
\end{equation}
which correspond to the simple idempotents $p^{(s,a) (s,a)}_A$ of the $Q$-algebra. The labels $\alpha$ in \eqref{eq:IdempotentDef} correspond to pairs $(s,a)$ of an input string $s\in \mathcal{C}$ and an integer $0\leq a \leq n_{A,s}$ for the $Q$-algebra [c.f. discussion above Eq. ~\eqref{eq:NonCommutative_Idempotents}]. Then $\ket{\psi}$ is an unit eigenvector of $\Pi_B^{(s,a)(s,a)}(p)$ if the excitation $B$ is of the $(s,a)$ subtype, while $\Pi_B^{(s,a)(s,a)}(p)\ket{\psi}=0$ otherwise:
\begin{equation}
    \Pi_{A}^{(s,a)(s^\prime,a^\prime)}(p)\ket{\psi} = \delta_{A,B}\delta_{s,s^\prime}\delta_{a,a^\prime} \ket{\psi}.
    \label{eq:Idempotent_state}
\end{equation}

In particular, the operator $\Pi_{\mathbf{1}}^{11}(p)=B_p$ projects onto the local ground-state at plaquette $p$. The total number of subtypes of the anyon $B$ is given by $n^Q_B$, which is equivalent to the number of distinct states that realize a $B$ excitation at a given plaquette of the $LW_1$ model. 

These additional configurations that arise from the string-net construction should be taken into account in the computation of the Hilbert space dimension. In this case, the emergent topological order is described by $\mathcal{U}=\mathcal{Z}(\mathcal{C})$ and equation \eqref{eq:MSB} should be modified as follows:
\begin{equation}
\begin{split}
        \dim_{\mathcal{Z}(\mathcal{C})}\left(g,\{A_n\}_{n\leq m}\right) &= \sum_{C\in \mathcal{Z}(\mathcal{C})} S_{\textbf{1},C}^{2-2g-m} \prod_{n=1}^m S_{A_n,C}n^Q_{A_n}.\\
\end{split}
    \label{eq:FullHSDim_LW1}
\end{equation}

Given the analytical expression \eqref{eq:FullHSDim_LW1}, it becomes trivial to compute the spectral degeneracies for the $LW_1$ model. Notice that using the Hamiltonian \eqref{eq:Hamiltonian}, with appropriate $B_p$ operators, the energy of a state containing $m$  non-trivial $\left( A_n \neq \mathbf{1}\right)$ excitations is $E_0 + m$, with the ground state energy being $E_0=-N_p$. To obtain an expression for the degeneracy of the $m-\rm{th}$ excited state, $D(g,m)$, one must (i) enforce that the anyons $A_n \in \mathcal{Z}(\mathcal{C})^* \equiv \mathcal{Z}(\mathcal{C})\backslash \{\mathbf{1}\}$ are non-trivial (ii) sum equation \eqref{eq:FullHSDim_LW1} over all possible labels of each $A_n$  and (iii) include a combinatorial prefactor that accounts for choosing which plaquettes of the graph host the excitations. We obtain:

\begin{equation}
\begin{split}
        D(g,m) &= \binom{N_p}{m}\sum_{A_1,\dots,A_m \in \mathcal{Z}(\mathcal{C})^*}    \dim_{\mathcal{Z}(\mathcal{C})}\left(g,\{A_n\}_{n\leq m}\right)\\
        &= \binom{N_p}{m} \sum_{C\in \mathcal{Z}(\mathcal{C})} S_{\textbf{1},C}^{2-2g} \left( \frac{\left[S \mathbf{n}^Q\right]_C}{S_{\mathbf{1},C}}-1 \right)^m,
\end{split}
\label{eq:FullDegeneracy_LW1}
\end{equation}
where we have defined a \emph{multiplicity vector} $\mathbf{n}^Q$,  of dimension equal to the number of simple objects in $\mathcal{Z}(\mathcal{C})$ and entries equal to $[\mathbf{n}^Q]_A\equiv n^Q_A$, and introduced the notation
\begin{equation}
    \left[S \mathbf{n}^Q\right]_C = \sum_{A \in \dc} S_{C,A}n^Q_{A},
\end{equation}
which is viewed as a matrix multiplication of the $S$-matrix and the multiplicity vector. This is a main result of this paper, serving as a basis for the computation of partition functions of all extended $LW_k$ models.

Note that by replacing $\mathbf{n}^Q$ by $\mathbf{n}^O$ with $n^O_A = n^Q_{A,1}$ and using the identity $S\mathbf{n}^O=\mathbf{n}^O$ in Eq.~\eqref{eq:FullDegeneracy_LW1} one recovers the results for the $LW_0$ model~\cite{Ritz-Zwilling2024_2}.

\subsection{\label{subsec:PartitionFunctions} Partition function for the \texorpdfstring{$LW_1$}{Lg} model}
To set up the study of the finite-temperature properties of the extended string-net model $LW_1$, which contains contributions from all the anyons of the output topological order, we determine the partition function $Z(\beta,g,N_p)$. Since we have an exact expression for the degeneracy of each energy level $E_m$, this function follows directly:
\begin{equation}
    \begin{split}
        Z(\beta,g,N_p) &= \sum_m D(g,m)e^{-\beta E_m}\\
         &= \sum_{C\in \mathcal{Z}(\mathcal{C})} S_{\textbf{1},C}^{2-2g} \left( \frac{\left[S \mathbf{n}^Q\right]_C}{S_{\mathbf{1},C}}-1 +e^{\beta} \right)^{N_p}.
    \end{split}
    \label{eq:PartitionFunction_LW1}
 \end{equation}

Here as well, replacing $\mathbf{n}^Q$ by $\mathbf{n}^O$, one recovers the results for the $LW_0$ model~\cite{Ritz-Zwilling2024_2}:
\begin{equation}
        Z(\beta,g,N_p) 
         = \sum_{C\in \mathcal{Z}(\mathcal{C})} S_{\textbf{1},C}^{2-2g} \left( \frac{n_{C,1}^Q}{S_{\mathbf{1},C}}-1 +e^{\beta} \right)^{N_p}.
    \label{eq:PartitionFunction_LW0}
 \end{equation}

\subsubsection{\label{subsubsec:Punctures} Surfaces with holes}

There are many ways to introduce gapped boundaries in string-net models \cite{Kitaev2012,Lan2015} which are compatible with the achiral topological orders emerging in these systems. Here we consider holes on the underlying orientable manifold where the string-net graph is embedded.  A hole is considered as a special kind of plaquette, that is not summed over in the Hamiltonian, and that is surrounded by a number of edges that may be different from that of a bulk plaquette.  A subtlety that emerges in the extended string-net models is that these holes may contain tails. Here, we consider holes that have the same number of tails attached as the plaquettes in the bulk lattice,  see Fig. \ref{fig:Puncture_Lattice} for the example of $LW_1$. In this way,  holes may host anyons through the same mechanism as detailed above, but without incurring any energetic cost.

\begin{figure}
    \centering
    \resizebox{!}{7cm}{\input{Puncutre_Lattice.tikz}}
    \caption{Realization of a hole (marked with blue horizontal stripes) on the $LW_1$ extended string-net model's lattice, with the edges and tail of the hole colored in red. The position of the tail on the hole is arbitrary.}
    \label{fig:Puncture_Lattice}
\end{figure}

To count the degeneracies in this case, we have to compute the number of fusion trees like the one depicted in Fig.~\ref{fig:MSB_Fusion}, with the addition of an extra branch containing the quasiparticles hosted in the holes and including the respective internal degeneracies. For a system with $b$ holes each hosting an anyon $B_j 
\in \dc, j\leq b$, the Hilbert space dimension associated to configurations with a set of $m$ excitations $\{A_n\}$ is given by: 
\begin{equation}
\begin{split}
  \dim(g,\{A_n\},b,\{B_j\}) = &\sum_{C\in \mathcal{Z}(\mathcal{C})} S_{\textbf{1},C}^{2-2g-m} \times  \\
  & \times \prod_{n=1}^m S_{A_n,C}n^Q_{A_n} \prod_{j=1}^b S_{B_j,C}n^Q_{B_j}.
\end{split}
  \label{eq:HSDimBoundaries}
\end{equation}

Since the anyons $B_j$ in the holes do not cost energy, the corresponding partition function can be computed as:
\begin{equation}
\begin{split}
    Z(\beta,g,N_p)  = \sum_{C\in \mathcal{Z}(\mathcal{C})} S_{\textbf{1},C}^{2-2g}& \left( \frac{\left[S \mathbf{n}^Q\right]_C}{S_{\mathbf{1},C}}-1 +e^{\beta} \right)^{N_p} \\
    \times & \left( \frac{\left[S \mathbf{n}^Q\right]_C}{S_{\mathbf{1},C}} \right)^{b}.
\end{split}
    \label{eq:BoundaryPartition}
\end{equation}

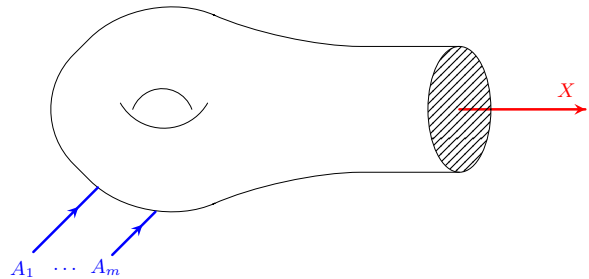
\begin{figure}
    \centering
    \resizebox{!}{4cm}{\begin{tikzpicture}

\begin{scope}[line cap=round,decoration={
	markings,
	mark=at position 0.7 with {\arrow{stealth}}}]

    \draw[D1, rounded corners=35pt](7,-1)--(4.2,-1)--(2,-2)--(0,0) -- (2,2)--(4.2,1)--(7,1);

    \draw[blue, postaction=decorate, line width=0.4mm] ($(-135:2.5) + (2,-0.5)$)--++(45:1.45);
    \draw[blue, postaction=decorate, line width=0.4mm] ($(-135:2) + (2,-0.5) + (0.9,-0.4)$)--++(45:.97);

    \node[blue] at ($(-135:2.6) + (2,-0.5) + (-0.1,-0.2)$) {$A_1$};
    \node[blue] at ($(-135:2.6) + (2,-0.5) + (-0.1,-0.2) + (0.7,0)$) {$\dots$};
    \node[blue] at ($(-135:2) + (2,-0.5) + (0.9,-0.4)+ (-0.1,-0.2)$) {$A_m$};
    
    \draw[D1] (2.75,0.0) arc (20:160:0.5 and 0.5);
    \draw[D1] (3.,0.1) arc (-30:-150:0.8 and 0.8);

\end{scope}

\begin{scope}[line cap=round,decoration={
	markings,
	mark=at position 1 with {\arrow{stealth}}}]

     \draw[red, line width= 0.4mm, postaction=decorate] (7,0)--++(2,0);
    \node[red] at ($(8.7,0.3)$) {$X$};
    \draw[D1, pattern=north east lines] (7.5,0) arc (0:360:0.5cm and 1cm);
\end{scope}

\end{tikzpicture}}
    \caption{A manifold of genus $g=1$ with a hole (shaded disk). A fixed quantum number $X\in \dc$ is penetrating the hole and quasiparticles $A_n$ hosted at the plaquettes on the surface are also represented.}
    \label{fig:ManifoldX}
\end{figure}

Holes of two different manifolds  $\mathcal{M}_1$ and $\mathcal{M}_2$ can be sewn together, an operation known as a connected sum (for simplicity, we assume they each have only one hole). The resulting manifold $\mathcal{M}=\mathcal{M}_1 \cup \mathcal{M}_2$ has genus $g=g^{\mathcal{M}_1}+g^{\mathcal{M}_2}$ and $N_p = N_p^{\mathcal{M}_1} + N_p^{\mathcal{M}_2} $ plaquettes.

Because each of the initial manifolds is orientable, this connected sum is unique and only possible if the quantum number $X\in \dc$ exiting $\mathcal{M}_1$ is the same as the one entering into $\mathcal{M}_2$ (i.e. if it is equal to $\bar{X}$) \cite{Ritz-Zwilling2024_2, Simon2023}. It is useful to define effective partition functions $Z_X$, where the quantum number at the hole is fixed to $X$, as shown in Fig.~\ref{fig:ManifoldX}:
\begin{equation}
    Z_X(\beta, g, N_p^{\mathcal{M}}) = \sum_{C \in \dc} S_{\bar{X},C} S_{\mathbf{1},C}^{1-2g} \left( \frac{\left[S \mathbf{n}^Q\right]_C}{S_{\mathbf{1},C}}-1 +e^{\beta} \right)^{ N_p^{\mathcal{M}}}.
    \label{eq:EffectivePartitionFunction}
\end{equation}

The partition function for the resulting manifold of the connected sum of $\mathcal{M}_1$ with $\mathcal{M}_2$ is simply:
\begin{equation}
    Z(\beta,g,N_p) = \sum_{X\in \dc} Z_X(\beta,g^{\mathcal{M}_1},N_p^{\mathcal{M}_1})Z_{\bar{X}}(\beta,g^{\mathcal{M}_2},N_p^{\mathcal{M}_2}).
\end{equation}

\subsubsection{\label{subsec:subset} Subset of fluxons \texorpdfstring{$\mathcal{N}$}{Lg}}
The quantity $S\mathbf{n}^Q$ inside the brackets of Eq.~\eqref{eq:PartitionFunction_LW1} plays a central role in understanding what are the contributions of each term to the partition function. We call $\mathcal{N}$ the subset of anyons $A$ that are such that $[S \mathbf{n}^Q]_A\neq 0$. It can be shown that:
\be
\left[S \mathbf{n}^Q\right]_A = 0 \text{ if } A \notin \mathcal{F},
\label{eq:NisFluxonx}
\ee
which implies that $\mathcal{N}\subseteq \mathcal{F}$. Also, for the entry corresponding to the vacuum term $A=\mathbf{1}$, it can be analytically found that:
\begin{equation}
    \left[S \mathbf{n}^Q\right]_\mathbf{1} = \sum_{s\in \mathcal{C}} d_s.
    \label{eq:SN_1}
\end{equation}

Note the difference with the internal multiplicities of the $O$-algebra (with no tails) that satisfy
\begin{equation}
\left[S \mathbf{n}^O\right]_\mathbf{1} = n^O_\mathbf{1} = 1 .
\label{eq:Sn^0}
\end{equation}

We can also show that if $\mathcal{C}$ is Abelian then $\mathcal{N}=\{\mathbf{1}\}$, such that:
\be
[S\mathbf{n}^Q]_A =\mathcal{D }\delta_{A,\mathbf{1}}.
\label{eq:SN_Abelian}
\ee

Details on the derivation of Eqs.~\eqref{eq:NisFluxonx}-\eqref{eq:SN_Abelian} are given in Appendix \ref{appendix:SpecialSubset}. For a non-Abelian $\mathcal{C}$, we are not able to make a general proof, but numerical studies of $S\mathbf{n}^Q$ for several input UFCs lead us to conjecture that $\mathcal{N}=\{\mathbf{1}, B\}$, where $B\in \mathcal{F}^*$ is a non-trivial fluxon such that:
\be
[S\mathbf{n}^Q]_\mathbf{1} = \sum_{s\in \mathcal{C}} d_s > [S\mathbf{n}^Q]_B /d_B >0.
\ee
The identification of $B$ is non-trivial and depends on the structure of the tube algebra's tensors $M^{11}_B$. In Appendix~\ref{appendix:SpecialSubset}, we obtain a closed expression for the components $[S\mathbf{n}^Q]_A$ in terms of the tube algebra's tensors.

\subsubsection{\label{Thermodynamical limit} Thermodynamic limit}
A consequence of the above results is that the term in brackets of Eq.~\eqref{eq:PartitionFunction_LW1} is always maximized by the vacuum term, i.e. $[S\mathbf{n}^Q]_\mathbf{1} > [S\mathbf{n}^Q]_C/d_C, \forall C\neq \mathbf{1}$. Thus, it dominates the behavior of the partition function in the thermodynamical limit $N_p\rightarrow \infty$:
\begin{equation}
  \lim_{N_p \rightarrow \infty} Z(\beta,g, N_p)  \sim \mathcal{D}^{2-2g}  \left( \mathcal{D} \sum_{s\in \mathcal{C}} d_s - 1 + e^\beta \right)^{N_p},
   \label{eq:ThermodynamicalLimit}
\end{equation}
where $\mathcal{D}=\mathcal{D}_{\mathcal{C}}^2$ is the total quantum dimension of $\dc$. Thus the dominant term of the partition function can be obtained with only the knowledge of the input data, that is, the genus $g$ and ``size" $N_p$ of the input graph alongside the input UFC $\mathcal{C}$.

In Eq.~\eqref{eq:ThermodynamicalLimit}, we recognize the partition function 
\begin{equation}
(q-1+e^{\beta})^{N_p},
\end{equation} 
for $N_p$ independent spins with 
\begin{equation}
q\equiv \mathcal{D} \sum_{s\in \mathcal{C}} d_s,
\label{eq:qdef}
\end{equation} 
states each.

In Ref.~\cite{Ritz-Zwilling2024_2}, the partition function for the $LW_0$ model was computed, yielding a structure very similar to equation \eqref{eq:PartitionFunction_LW1}, but with a multiplicity vector with components $[\mathbf{n}^O]_A = n^Q_{A,1}$, that are non-zero only for fluxons, i.e. $A\in \mathcal{F}$. In this case, a subset of fluxons (called \emph{pure fluxons}) $\mathcal{P}$ was identified to maximize the term in brackets of the corresponding partition function, i.e. $A\in \mathcal{P}$ iff $n^Q_{A,1}=d_A$~\cite{Ritz-Zwilling2024_2}. However, here we see that once all anyons in $\dc$ are allowed to occupy the plaquettes of the model, then this ``degeneracy" is lifted and only the vacuum $\mathbf{1}$ contribution is dominant in the thermodynamic limit.

\subsubsection{\label{GDS_HSDim} Ground-state degeneracy and Hilbert space dimension }

Before proceeding with the study of finite temperature properties, we consider the partition function in some interesting limiting cases. One feature of topological phases is that the ground-state degeneracy (GSD) depends on the topology of the surface the system lives on. Indeed, one can compute the GSD from the partition function \eqref{eq:PartitionFunction_LW1} in the zero temperature limit:
\begin{equation}
    \text{GSD} = \lim_{\beta \rightarrow \infty} Z(\beta,g,N_p)e^{\beta E_0}=\sum_{C\in \mathcal{Z}(\mathcal{C})} S_{\textbf{1},C}^{2-2g},
    \label{eq:GSD}
\end{equation}
which is precisely the result obtained for the GSD of a TQFT \cite{Moore_1989}.

The Hilbert space dimension is obtained in the opposite limit $\beta \to 0$:
\begin{equation}
    \text{dim }\mathcal{H} = \lim_{\beta \rightarrow 0} Z(\beta,g,N_p)=\sum_{C\in \mathcal{Z}(\mathcal{C})} S_{\textbf{1},C}^{-N_v/2} \left[S \mathbf{n}^Q\right]_C^{N_p},
    \label{eq:dimH}
\end{equation}
where $N_v=2(N_p-2+2g)$ is the number of vertices for a trivalent graph on a closed surface of genus $g$.

\subsection{\label{subsec:Thermodynamics} Specific heat}

Picking up on equation \eqref{eq:PartitionFunction_LW1}, it is straightforward to compute finite-temperature properties. In the thermodynamic limit, we may use the simpler expression \eqref{eq:ThermodynamicalLimit} to compute the specific heat:
\begin{equation}
        c = \lim_{N_p \rightarrow \infty} \, -\frac{\beta^2}{N_p}    \dfrac{\partial^2 Z}{\partial \beta^2} =   \,\frac{\beta^2 e^\beta \left(q - 1 \right)}{\left(q - 1 + e^\beta \right)^2}.
        \label{eq:SpecificHeat}
\end{equation}

This expression holds for any input UFC and has the same structure as the results found in \cite{Ritz-Zwilling2024_2,Vidal2022}. Since $c$ is a smooth function of the temperature, not displaying any divergences nor discontinuities, it shows that there is no finite-temperature phase transition in the $LW_1$ model. As it will become clear in the next section, this conclusion also holds for any of the other models $LW_k$, which is in accordance with the more general result derived by Hastings \cite{Hastings2011} that states the absence of such a transition for any  Hamiltonian that is a sum of local commuting projectors. This form of the specific heat also displays a maximum at low temperatures, known as a Schottky anomaly. It usually arises in systems with few discrete energy levels, such as independent spins with $q$ states or the 1D classical Potts model with $q$ colors.

\subsection{\label{subsec:Generalizing_To_LW_k} Generalizing to \texorpdfstring{$LW_k$}{Lg}}

In the previous sections, we have been working with the $LW_1$ model. Nonetheless, it is a trivial matter to generalize the results to the rest of the family of extended string-net models $LW_k$, given the considerations of section~\ref{subsec:MoritaEquivalence}.

The excitations arising in all of these models are described by the same UMTC $\dc$ for a given input UFC $\mathcal{C}$, but the internal multiplicities $n_A^{(k)}$ are different. This means that on a plaquette with $k$ tails, an anyon $A$ can appear with $n_A^{(k)}$ distinct subtypes. Incorporating this change in the description of the previous sections, we define a generalized multiplicity vector $\mathbf{n}^{(k)}$ with $|\dc|$ components $[\mathbf{n}^{(k)}]_A=n_A^{(k)}$ (and $\mathbf{n}^{(1)}=\mathbf{n}^Q$) such that Eq.~\eqref{eq:FullDegeneracy_LW1} should now read
\begin{equation}
     D(g,m) = \binom{N_p}{m} \sum_{C\in \mathcal{Z}(\mathcal{C})} S_{\textbf{1},C}^{2-2g} \left( q_C^{(k)} -1 \right)^m
     \label{eq:FullDegeneracy_LWk}
\end{equation}
and the partition function
\begin{equation}
      Z(\beta,g,N_p) = \sum_{C\in \mathcal{Z}(\mathcal{C})} S_{\textbf{1},C}^{2-2g} \left( q_C^{(k)}-1 +e^{\beta} \right)^{N_p},
\label{eq:PartitionFunction_LWk}
\end{equation}
where we introduced the notation:
\begin{equation}
    q_C^{(k)}=\frac{\left[S \mathbf{n}^{(k)}\right]_C}{S_{\mathbf{1},C}}.
\label{eq:qCk}
\end{equation}

An important subtlety, already noticed for the Vec($\mathbb{Z}_2$) input UFC in \cite{MapMorita}, is that $P_\mathbf{1}\neq B_p$ for $k\geq 2$, meaning that the vacuum $\mathbf{1}\in \dc$ has more than one subtype $[\mathbf{n}^{(k)}]_\mathbf{1} \geq 2$ . One subtype is always given by the ``Kirby strand" \cite{Simon2023} which is associated to the plaquette operator $B_p$, such that it costs no energy. In contrast, the other subtypes are associated with non-trivial tails that, when combined, fuse in the vacuum, but which induce an energy cost nonetheless. Thus, when summing over the non-trivial anyons $A_n \in \dc^*$ to compute the degeneracies of excited states, the set $\dc^*$ must be redefined to contain the ``false vacuums" that have an energy cost, while excluding the subtype associated to the ``true vacuum". In the following sections, when considering the $LW_1$ model we use $\dc^*$ and $\dc/\{\mathbf{1}\}$ interchangeably, and in section \ref{subsec:Fibonacci} we further explore the distinction between the true and false vacuums for a particular UFC.

In principle, one could use the relation in Eq.~\eqref{eq:Nk_from_N1} to compute the quantity $S\mathbf{n}^{(k)}$ from  $S\mathbf{n}^Q$ and the fusion matrices $N_a$. However, when the input UFC $\mathcal{C}$ is Abelian, it can be shown that:
\begin{equation}
    \left[ S\mathbf{n}^{(k)}\right]_C = \left(\left[ S\mathbf{n}^Q\right]_C\right)^k.
    \label{eq:Snk}
\end{equation}

Interestingly, we numerically find that this identity seems to hold for non-Abelian UFCs as well, although it is difficult to prove this statement in the general case. From this property, it is clear that the vacuum anyon is still the dominant contribution in the partition function in the thermodynamic limit [with $q_\mathbf{1}^{(k)}= \mathcal{D} (\sum_{s\in \mathcal{C}} d_s)^k$] and that the subset of fluxons $\mathcal{N}$ still plays a central role at finite size. This property also indicates that models with a higher number of tails converge more quickly to the thermodynamic limit.

\section{\label{sec:finitesize}Non-trivial scaling between temperature and system size}

It is well known that the topological orders arising in string-net and quantum double models in 2D are destroyed at any non-zero temperature in the thermodynamic limit~\cite{Castelnovo2007,Nussinov2008,Hastings2011}. However, for finite-size systems, these orders may survive up to some finite temperature.  In this section, we study corrections to the thermodynamic limit of the partition function and compute the thermal averages of different physical observables in a  finite-size system: topological projectors, Wegner-Wilson loops and topological mutual information. All these quantities reveal a non-trivial scaling between temperature and system size.

\subsection{\label{subsec:Partition_Corrections} Corrections to the partition function}
The partition function in Eq.~\eqref{eq:PartitionFunction_LW1} is dominated by the vacuum $\mathbf{1}$ in the thermodynamic limit, and corrections to this limit are given by the second-largest term. Due to the similarity between \eqref{eq:PartitionFunction_LW1} and the partition function of classical spin chains, such as the Potts model~\cite{Kardar_book}, we can extract an effective ``correlation area" for the $LW_1$ model by examining the ratio between the leading term and these corrections~\cite{Ritz-Zwilling2024_2}.

As stated in the previous section for non-Abelian UFCs, there are only two fluxons $A\in \mathcal{N}=\{\mathbf{1},B\}$ for which $[S\mathbf{n}^Q]_A\neq 0$, while for Abelian UFCs there is only the vacuum $\mathcal{N}=\{\mathbf{1}\}$. Let $\mathcal{B}$ denote the set of anyons $A$ that have the second largest 
\begin{equation}
q_A=\frac{[S\mathbf{n}^Q]_A}{S_{\mathbf{1},A}}=\mathcal{D}\frac{[S\mathbf{n}^Q]_A}{d_A},
\end{equation}
after the vacuum $q_\mathbf{1}=[S\mathbf{n}^Q]_\mathbf{1} \mathcal{D}=q$  [see Eq.~\eqref{eq:qdef}]. Then $\mathcal{B}=\{B\}$ for non-Abelian (resp. $\mathcal{B}=\dc^*$ for Abelian) input UFCs with $q_A>0$ (resp. $q_A=0$) for any $A\in \mathcal{B}$. Therefore, the partition function reads
\begin{equation}
    Z(\beta,g,N_p) \approx Z_{\infty} \left(1 + \frac{\sum_{A \in \mathcal{B}} S_{1,A}^{2-2g}}{S_{\mathbf{1,\mathbf{1}}}^{2-2g}} e^{-\frac{N_p}{N_p^*}} \right),
    \label{eq:PartitionCorrections}
\end{equation}
where $Z_{\infty}$ represents the partition function in the thermodynamic limit. This equation defines the correlation area $N_p^*$:

\begin{equation}
\begin{split}
     N_{p}^* &= \left[ \log\left(\frac{q -1+e^\beta}{q_A-1+e^\beta}\right) \right]^{-1}\\
     &\underset{\beta \rightarrow \infty}{\sim} \frac{e^\beta}{q-q_A},
\end{split}
\label{eq:Np_Star1}
\end{equation}
where we may take any $A \in \mathcal{B}$. This area diverges in the low-temperature limit, which again signals the presence of a phase transition at a critical temperature $T_c = 0^+$. This transition separates a phase with topological order at $T=0$ with a phase at $T>0$ with no topological order.

The correlation area can also be computed for the more general $LW_k$ model:

\begin{equation}
     N_{p}^{(k)*} = \left[ \log\left(\frac{q^{(k)} -1+e^\beta}{ q_A^{(k)}-1+e^\beta}\right) \right]^{-1}.
     \label{eq:NpStark}
\end{equation}

It amounts to substitute the parameters $q/q_A$ by their analogues $q^{(k)}/q^{(k)}_A$ [see Eq. ~\eqref{eq:qCk}]. By virtue of Eq. ~\eqref{eq:Snk}, and given that $q^{(k)} > q^{(k)}_A$ for any $A \in \mathcal{B}$, this correlation area decreases as the number $k$ of tails per plaquette increases. In the limit of a large number of tails $k\rightarrow \infty$, the correlation area approaches zero, and the partition function matches the thermodynamic limit $Z_{\infty}$.

\subsection{\label{Thermal_Average_of_Projectors} Topological projectors}

\begin{figure}
    \centering
         \resizebox{8cm}{!}{\begin{tikzpicture}[thick,decoration={
		markings,
		mark=at position 1.1 with {\arrow{stealth}}},
		baseline={([yshift=-.5ex]current bounding box.center)},
		line cap=round
	] 
	
\begin{scope}[black] 
\draw[D1, line width=0.4mm] (-135:1.5)--++ (45:0.6);

\draw[-{Stealth[length=5mm]},D1, line width=0.4mm] (-135:0.6)-- node[above=4mm,right=7mm] {\large{$B$}}  ($(-135:0.7) + (45:2.5)$);
 \end{scope}

\node (a) at (-3,0) {};
\node (b) at (3,0) {};
\node (c) at ($(a)!0.5!(b)$) {};

\begin{scope}[shift={(c)},x={(a)}, scale=0.7]

\draw[blue, line width=0.8mm, postaction=decorate] (0.3,0) arc  (0:130:0.3 and 1.5) node[above=5mm,left=5mm] {\large{$P_A(L_c)$}};
\draw[blue, line width=0.8mm] (0.3,0) arc (0:-200:0.3 and 1.5);
\end{scope}

\node (d) at (1.7,0) {\huge{$=$}};
\node (e) at (2.8,0) {\huge{$\delta_{A,B}$}};

\draw[-{Stealth[length=5mm]},D1, line width=0.4mm] ($(4,0) + (-135:1.5)$) -- node[above=4mm,right=9mm] {\large{$B$}}   ($(4,0) + (-135:1.5) + (45:3.2)$);

\end{tikzpicture}}    
    \caption{Schematic representation of the topological projectors $P_A(L)$}
    \label{fig:Topological_Projectors}
\end{figure}
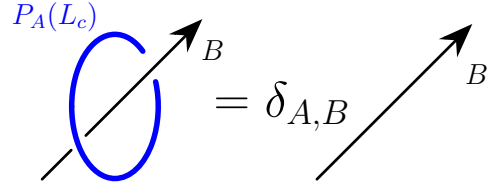

Following \cite{Ritz-Zwilling2024_2}, we can derive interesting scaling behaviors by studying the thermal properties of topological projectors $P_A(L)$, for a loop (closed path) $L$ and $A\in \dc$. These operators can be interpreted as projectors onto states where all anyons inside the loop $L$ fuse into $A$, as is schematically shown in Fig. \ref{fig:Topological_Projectors}. They form a complete set of orthogonal projectors, in the sense that they satisfy:
\begin{equation}
    P_A(L)P_B(L) = \delta_{A,B}P_A(L) \quad \quad \sum_{A \in \dc} P_A(L) = \mathds{1}
    \label{eq:Projectors_Relations}
\end{equation}

Here we will only consider contractible loops $L$, which may be characterized by an outer region $\mathcal{R}$, with $N_p^{\mathcal{R}}$ plaquettes and genus $g^{\mathcal{R}}=g$ and the inner (complementary) region $\bar{\mathcal{R}}$ with $N_p^{\bar{\mathcal{R}}}=N_p-N_p^\mathcal{R}$ plaquettes and genus $g^{\bar{\mathcal{R}}}=0$  (necessary for the loop to be contractible).   The thermal average of these projectors can be computed using the effective partition functions $Z_X(g,\beta)$ defined in Eq.~\eqref{eq:EffectivePartitionFunction}  for surfaces with quantum number $X \in \dc$ exiting through a puncture introduced in section \ref{subsubsec:Punctures}:
\begin{equation}
    \langle P_X(L) \rangle = \frac{Z_X(g,\beta, N_p^{\mathcal{R}})Z_{\bar{X}}(0,\beta, N_p^{\bar{\mathcal{R}}})}{Z_X(g,\beta,N_p)} .
    \label{eq:ThermalAverageProjector}
\end{equation}

Although this expression is structurally complex for an arbitrary loop $L$, it reduces to a much simpler form in the limit where the outer region is very large:

\begin{equation}
\begin{split}
    \lim_{N_p^{\mathcal{R}}\rightarrow \infty}   \langle P_X(L) \rangle  = & d_X \sum_{C \in \dc} S_{\mathbf{1},C}S_{\bar{X},C} \\
    \times & \left( \frac{q_C - 1 + e^\beta}{q - 1 + e^\beta}\right)^{N_p^{\bar{\mathcal{R}}}} 
\end{split}
    \label{eq:Projector_ThermodynamicLimit}
\end{equation}

From this expression it is trivial to check that in the limit of zero temperature:
\begin{equation}
    \lim_{T\rightarrow 0} \langle P_X (L) \rangle = \delta_{X,\mathbf{1}},
    \label{eq:ThermalProjectort_GS}
\end{equation}
which yields the expected result that, for all ground states of string-net models, contractible loops cannot contain non-trivial quasiparticles.

\subsection{\label{subsubsec:WWL} Wegner-Wilson loops}
Wegner-Wilson loops (WWL) are closed-string operators $W_A(L)$ directly related to the topological projectors $P_A(L)$ that can be defined through the unlinking relation \cite{Kitaev2006, Simon2023}: 
\begin{equation}
    W_A(L) = \sum_{B} \frac{S_{A,B}}{S_{\mathbf{1},B}} P_B(L).
    \label{eq:WWL_Def}
\end{equation}
The operator $W_A(L)$ is defined along a contractible loop $L$ encircling a region $\bar{\mathcal{R}}$. 

At zero temperature, the expectation value $\langle W_A(L) \rangle$ allows one to study the (de)confinement of the anyon $A$ as a function of loop size~\cite{RitzZwilling2021,Schotte2019,MoessnerMoore}. If  $\langle W_A(L) \rangle$ follows a ``perimeter law'', i.e. if it decays exponentially with the perimeter $|L|$ of the loop (i.e. a slow decay), then $A$ is deconfined. This is also the case if $\langle W_A(L) \rangle$ does not decay at all, a behavior known as a ``zero law". Conversely, if $\langle W_A(L) \rangle$ follows an ``area law'', meaning that it decays exponentially with the area encircled by the loop $L$ (i.e. a faster decay), then the anyon is confined. We may write these relations as:
\begin{equation}
    \langle W_A(L) \rangle \sim 
    \begin{cases}
     e^{-c_1 |L|}, & \, A \text{ is deconfined}\\
     e^{-c_2 N_p^{\bar{\mathcal{R}}}}, & \, A \text{ is confined}
    \end{cases}\quad,
    \label{eq:WWLDecay}
\end{equation}
where $c_1\geq 0$, $c_2>0$ and $N_p^{\bar{\mathcal{R}}}$ is the number of plaquettes inside the region $\bar{\mathcal{R}}$ encircled by the loop, which is a measure of its area. 

At finite temperature, if the thermal average $\langle W_A(L) \rangle$ decays following an area law, it indicates a thermal proliferation of anyons (topological defects) that destroy the topological order~\cite{MoessnerMoore}.

Inserting Eq.~\eqref{eq:Projector_ThermodynamicLimit} in Eq.~\eqref{eq:WWL_Def} we can compute  $\langle W_A(L) \rangle$ for finite loops in the thermodynamic limit. We obtain:
\begin{equation}
    \begin{split}
        \lim_{N_p^{\bar{\mathcal{R}}} \rightarrow \infty} 
        \langle W_A(L) \rangle 
        &= d_A  \left( \frac{q_A - 1 + e^\beta}{q - 1 + e^\beta}\right)^{N_p^{\bar{\mathcal{R}}}}\\
&= d_A e^{-N_p^{\bar{\mathcal{R}}}/N_{A}^*(\beta)} \hfill,
   \end{split}
   \label{eq:WWLDecay_ThermodynamicLimit}
\end{equation}
with $N_{A}^*(\beta)$ being the characteristic ``decay area" for the anyon $A$ at inverse temperature $\beta$:

\begin{equation}
   N_A^{*}(\beta) = \left[ \log \left( \frac{q-1+e^\beta}{q_A-1+e^\beta} \right) \right]^{-1}.
    \label{eq:WWL_DecayArea}
\end{equation}

Notice the similarities with the correlation area $N_p^*$ defined in Eq.~\eqref{eq:Np_Star1}, although here $A \in \dc^*$. Indeed, we see that $N_A^*$ diverges as $e^{\beta}/(q-q_A)$ at low temperatures for all non-trivial anyons. This implies that all anyons follow a ``zero law" at $T=0$ and are deconfined, as is expected for a topologically ordered phase. For the vacuum $A=\mathbf{1}$, this remains true at all temperatures, while the other $A\in \dc^*$ follow an area law at any finite temperature.

For a given input category, most of the anyons satisfy $q_A=0$. In this case, the infinite temperature limit $\beta \to 0$ of Eq.~\eqref{eq:WWL_DecayArea} reveals that $N_A^*(\beta) \rightarrow 0$ and that these anyons are no longer present at any scale. However, for non-Abelian categories, there is one anyon $B$ such that $q_B > 0$ (cf. discussion in section \ref{Thermodynamical limit}) so that $N_B^*$ remains finite even in this limit. The anyon $B$ becomes relevant at scales smaller than:
\begin{equation}
    N_B^*(0) = \left[ \log \left( \frac{q}{q_B} \right) \right]^{-1}.
\end{equation}

\subsection{Topological mutual information}

To conclude this section, let us consider the behavior of topological mutual information (TMI) $I_\text{topo}$, usually introduced to probe topological orders at finite temperatures~\cite{Castelnovo2007a,Iblisdir2009,Iblisdir_2010}.

It is important to emphasize that TMI is different from the topological entanglement entropy (TEE)~\cite{Kitaev_TEE,Levin_2006}, defined as the constant term of the von Neumann entanglement entropy, which is commonly used to characterize topological orders at zero temperature. As pointed out in~\cite{Iblisdir2009,Wolf2008}, at finite temperatures the entanglement entropy does not measure only the entanglement between different regions, as it includes a volume term. The volume term cancels out in the definition of TMI, revealing contributions only from the topological order. The two quantities, TMI and TEE, coincide (up to a factor 2) at zero temperature.

\subsubsection{Finite-temperature conjecture}

Given a bipartition $(\mathcal{R},\bar{\mathcal{R}})$ of a system, with common boundary $L$, the mutual information is defined as:

\begin{equation}
    I_{\mathcal{R}}=I_{\bar{\mathcal{R}}} = S_{\mathcal{R}} + S_{\bar{\mathcal{R}}} - S_{\mathcal{R} \cup \bar{\mathcal{R}}},
    \label{eq:MutualInfo}
\end{equation}
with $S_{\mathcal{R}} = -\text{Tr} (\rho_\mathcal{R} \log (\rho_\mathcal{R}))$ the von Neumann entropy and $\rho_\mathcal{R} = \text{Tr}_{\bar{\mathcal{R}}}(\exp (-\beta H))/Z$. In the limit $|L|\rightarrow \infty$, the mutual information behaves as
\begin{equation}
    I_\mathcal{R} = \alpha |L| - \gamma,
\end{equation}
which is often called an ``area law", although $|L|$ represents the perimeter of the loop in this 2D setup. The constant term $-\gamma$ is the definition of TMI, thus $I_{\text{topo}}=-\gamma$. Under general assumptions, Iblisdir \emph{et al.} \cite{Iblisdir2009,Iblisdir_2010} conjectured that at finite temperature, for the Kitaev quantum double model \cite{Kitaev_2003} on a surface with genus $g=0$, the TMI takes the following form
\begin{equation}
    I_{\text{topo}}(\beta) = -\sum_{A \in \dc} \langle P_A(L) \rangle \log \left( \langle P_A(L) \rangle \left[\frac{\mathcal{D}}{d_A}\right]^2 \right),
    \label{eq:ItopoConjecture}
\end{equation}
where $\langle P_A(L) \rangle$ is the thermal average of the topological projectors introduced in the beginning of the section. Since this applies to surfaces with $g=0$, the boundary loop of the bipartition $L$ is always contractible. Although this conjecture was built on a different context that targeted the quantum double models, we assume that is holds for the extended string-net models studied in this work.

\subsubsection{\label{subsubsec:Itopo_Scaling} Temperature and system size dependence}

In the limit of zero temperature, we can use Eq.~\eqref{eq:ThermalProjectort_GS} to evaluate the TMI, obtaining
\begin{equation}
    I_{\text{topo}}(T=0) = -2\log \mathcal{D},
    \label{eq:ZeroTemp_TMI}
\end{equation}
which coincides (up to a factor of 2) with the well known result for the TEE for a topological phase with total quantum dimension $\mathcal{D}$~\cite{Levin_2006,Kitaev2006}. Conversely, in the thermodynamic limit with the boundary $|L|\rightarrow \infty$, inserting Eq.~\eqref{eq:Projector_ThermodynamicLimit} into Eq.~\eqref{eq:ItopoConjecture} at any finite temperature yields:

\begin{equation}
    \lim_{N_p^\mathcal{R},N_p^{\bar{\mathcal{R}}}\rightarrow \infty} I_{\text{topo}}(T> 0) = 0.
    \label{eq:TMI_ThermodynamicLimit}
\end{equation}

These results are again in agreement with the emergence of a topologically ordered phase at $T=0$, which is destroyed at any finite temperature $T>0$ in the thermodynamic limit.

For finite-size systems, in order to inspect the scaling behavior of TMI we can study the corrections to the thermodynamic limit of the thermal average of topological projectors $\langle P_X(L)\rangle$. When $N_p^{\bar{\mathcal{R}}}$ is finite, the second highest contribution to the sum in Eq.~\eqref{eq:Projector_ThermodynamicLimit} is given by the anyons in the set $\mathcal{B}$ defined at the beginning of this section. If $B\in \mathcal{B}$ is a representative anyon of this set, then the first correction to the thermodynamic limit is given by:
\begin{equation}
    \langle P_X(L) \rangle \simeq \langle P_X(L) \rangle_{\infty} + \frac{d_X d_B}{\mathcal{D}} S_{\bar{X},B} e^{-N_p^{\bar{\mathcal{R}}}/N^*(\beta)},
    \label{eq:ThermalProjectors_Corrections}
\end{equation}
where $\langle P_X(L) \rangle_{\infty}$ is the thermal average in the thermodynamical limit and $N_p^*(\beta)$ is the characteristic ``decay area". Because the anyons $B \in \mathcal{B}$ maximize the quantity $q_B$, then:
\begin{equation}
     N_p^* = \max_{A \neq \mathbf{1}} N_A^*(\beta)
     =  \left[ \log \left( \frac{q-1+e^\beta}{q_B-1+e^\beta} \right) \right]^{-1},
    \label{eq:TMI_DecayArea}
\end{equation}
which is precisely the same quantity obtained in Eq.~\eqref{eq:Np_Star1}. When $N_p^{\bar{\mathcal{R}}}\gg N_p^*$ we recover the thermodynamic limit with $I_\text{topo}=0$ and in the low-temperature limit $N_p^*$ diverges, as given in Eq.~\eqref{eq:Np_Star1}. We can thus observe a crossover temperature $T_c$ when $N_p^{\bar{\mathcal{R}}}\simeq N_p^*$:
\begin{eqnarray}
    T_c &\simeq& \frac{1}{ \log [ N_p^{\bar{\mathcal{R}}}(q -q_B )]} \nonumber \\
    &\underset{ \text{large }N_p^{\bar{\mathcal{R}}}}{\sim}& \,\, \left(  \log  N_p^{\bar{\mathcal{R}}} \right)^{-1}
    \label{eq:TMI_CriticalTemperature}
\end{eqnarray}
This crossover temperature is similar to that of a classical Ising chain with $N_p^{\bar{\mathcal{R}}}$ spins.

\section{Example \label{sec:Examples}}

We will now illustrate the main results of the previous sections with the simplest non-Abelian input category, the Fibonacci UFC $\mathcal{C}=$ Fib. We will consider in detail the $k=0,1$ and $2$ cases of the family of extended string-net models, which are related to the $O$-, $Q$- and $\phi$-algebras respectively. For Fib, we also show how the multiplicity vector $\mathbf{n}^{(k)}$ can be obtained from the $Q$-algebra without the need to decompose the $k$-tails tube algebra.

\subsection{Fibonacci category and \texorpdfstring{$\mathcal{Z}$}{Lg}(Fib) \label{subsec:Fibonacci}}

The Fibonacci data consists of two self-dual string types $\{1, \tau\}$ and a single non-trivial fusion rule $\tau \times \tau = 1 + \tau$. Let $\varphi=(1+\sqrt{5})/2$ be the golden ratio that satisfies $\varphi^2 = \varphi+1$. Then the quantum dimensions of the input strings are $d_1=1$ and $d_\tau = \tau$. Incidentally, this category is modular and details on its $F$ and $R$ symbols can be found, for example, in \cite{NonAbelian_Review,Bonesteel2005}.

The emergent topological order of a string-net model with this input category is known as the doubled Fibonacci phase. There are 4 elements in the Drinfeld center $\mathcal{Z}$(Fib), usually denoted as $\{\mathbf{1}, 1\bar{\tau},\tau 1, \tau \bar{\tau} \}$ \footnote{Here the notation $\bar{\tau}$ should not be confused with the dual of $\tau$. It originated from the Drinfeld center $\mathcal{Z}(Fib) = Fib \otimes \overline{Fib}$ being a product of left and right-handed solutions of the hexagon equation} with quantum dimensions $\{1, \varphi, \varphi, \varphi^2 \}$ respectively, and an $S$-matrix given by:
\begin{equation*}
   S  =   \frac{1}{1+\varphi^2} \left(
\begin{matrix}
        1 & \varphi & \varphi & \varphi^2 \\
        \varphi & -1 & \varphi^2 & -\varphi \\
        \varphi & \varphi^2 & 1 & -\varphi \\
        \varphi^2 & -\varphi & -\varphi & 1
\end{matrix} \right) .
\end{equation*}

\subsection{\texorpdfstring{$Q$}{Lg}-algebra}

The decomposition of the $Q$-algebra in the doubled Fibonacci model is well established \cite{LanWen,Simon2023}. There is a total of seven tubes (here $e_1,e_2,\dots,e_7$ is a simple notation to denote the tubes, following \cite{LanWen}):

\begin{equation}
\begin{split}
    &e_1 = \mathcal{T}^{1;1}_{1;1} \quad \quad  e_2 = \mathcal{T}^{\tau;\tau}_{1;1} \quad \quad  e_3 = \mathcal{T}^{1;\tau}_{\tau;\tau} \quad \quad  e_4= \mathcal{T}^{\tau;1}_{\tau;\tau}\\
    &e_5 = \mathcal{T}^{\tau;\tau}_{\tau;\tau}  \quad \quad  e_6 = \mathcal{T}^{\tau;\tau}_{1;\tau} \quad \quad  e_7 = \mathcal{T}^{\tau;\tau}_{\tau;1}
\end{split}
\end{equation}

Tubes $e_6$ and $e_7$ belong to ``off-diagonal" sectors of the tube algebra (c.f. Appendix ~\ref{subAppendix:TAsectors}) and so they cannot be a part of the simple idempotents. The other five tubes can be arranged into five simple idempotents:

\begin{subequations}
\begin{align}
&p_{A}^{11} = \frac{1}{1+\varphi^2} (e_1 + \varphi e_2) \label{subeq:QSimpleA}\\
&p_{B}^{11} = \frac{1}{1+\varphi^2}(\varphi^2 e_1 - \varphi e_2)  \label{subeq:QSimpleB}\\
&p_{C}^{\tau\tau}=\frac{1}{1+\varphi^2}\left(e_3+\mathrm{e}^{-\frac{4 \pi \mathrm{i}}{5}} e_4+\sqrt{\varphi} \mathrm{e}^{\frac{3 \pi \mathrm{i}}{5}} e_5\right)  \label{subeq:QSimpleC} \\
& p_{D}^{\tau\tau}=\frac{1}{1+\varphi^2}\left(e_3+\mathrm{e}^{\frac{4 \pi \mathrm{i}}{5}} e_4+\sqrt{\varphi} \mathrm{e}^{-\frac{3 \pi \mathrm{i}}{5}} e_5\right)  \label{subeq:QSimpleD}\\
&p_{E}^{\tau\tau}=\frac{1}{1+\varphi^2}\left(\varphi e_3+ \varphi e_4+\sqrt{\varphi^{-1}} e_5\right)  \label{subeq:QSimpleE}
\end{align}
\label{eq:FibSimpleIdempotents}
\end{subequations}

Because $e_6 e_7 \propto p^{11}_B$ and $e_7e_6 \propto p_{E}^{\tau\tau}$, these two simple idempotents belong to the same 2-dimensional block of the tube algebra. Following the above notation for the elements of the Drinfeld center of $\mathcal{Z}$(Fib), the simple idempotents can be grouped into the following central idempotents:

\begin{subequations}
\begin{align}
      &P_{\boldsymbol{1}} = p^A_{11}  \label{subeq:QCentral1}\\
     &P_{1\bar{\tau}} = p^C_{\tau\tau} \label{subeq:QCentral1tau}\\
    &P_{\tau1} = p^D_{\tau\tau} \label{subeq:QCentraltau1}\\
     & P_{\tau\bar{\tau}} = p^B_{11} + p^E_{\tau\tau} \label{subeq:QCentraltautau}
\end{align}
\label{eq:FibCentralIdempotents}
\end{subequations}

Counting the number of simple idempotents that compose each central idempotent in \cref{subeq:QCentral1,subeq:QCentral1tau,subeq:QCentraltau1,subeq:QCentraltautau} yields their dimensions, which are used to construct the vector $\mathbf{n}^Q = (1,1,1,2 )^T$. Action by the $S$-matrix yields:
\begin{equation}
    S\mathbf{n}^Q= \left(\varphi^2,0,0,\varphi^{-2}\right)^T
\end{equation}
As explained in the sections above, the first entry (corresponding to the vacuum), also given by Eq.~\eqref{eq:SN_1}, has the highest value and dominates the partition function in the thermodynamic limit. The last term, corresponding to the fluxon $\tau \bar{\tau}$ dictates the crossover temperature and the decay/correlation area in finite-size systems. Thus the relevant subset of fluxons introduced in the previous sections are $\mathcal{N}=\mathcal{F}=\{\mathbf{1}, \tau \bar{\tau}\}$ and $\mathcal{B} = \{\tau \bar{\tau}\}$.

\subsection{\texorpdfstring{$O$}{Lg}-algebra}
The $O$-algebra can be obtained by restricting the $Q$-algebra to tubes containing only trivial tails. It is thus composed by the tubes $e_1$ and $e_2$ which correspond to closed loops of the input strings $1,\tau \in$ Fib. The two simple idempotents in this 2-dimensional $O$-algebra are given by Eqs.~\eqref{subeq:QSimpleA} and \eqref{subeq:QSimpleB}. In this case, these idempotents are also central, and correspond to the fluxons $\mathcal{F}=\{\mathbf{1}, \tau \bar{\tau}\}$ that satisfy $n_{F,1}\geq 1$:
\begin{subequations}
\begin{align}
&P_{\mathbf{1}} = \frac{1}{1+\varphi^2} (e_1 + \varphi e_2)\\
&P_{\tau \bar{\tau}} = \frac{1}{1+\varphi^2}(\varphi^2 e_1 - \varphi e_2)
\end{align}
\label{eq:OAlgebra_FibSimpleIdempotents}
\end{subequations}

The multiplicity vector is therefore $\mathbf{n}^O=(1,0,0,1)^T$.  As detailed in \cite{Ritz_Zwilling_2024}, the multiplicity vector of the $O$-algebra is an unit eigenvector of the $S$-matrix:
\begin{equation}
    S\mathbf{n}^O = \mathbf{n}^O
\end{equation}

\subsection{\texorpdfstring{$\phi$}{Lg}-algebra}
Tubes in the $\phi$-algebra have two tails on each side of the tube, as depicted in Fig. \ref{fig:Tube2}. For the Fib UFC there are a total of 47 tubes. 

We have constructed the Artin-Wedderburn decomposition of this tube algebra and found 13 simple idempotents (details of the decomposition and the tube algebra's tensors can be found the Appendix~\ref{appendix:TAdecomp}). They can be grouped into four central idempotents, corresponding to the objects in the Drinfeld center. One then obtains the multiplicity vector
\begin{equation}
    \mathbf{n}^\phi=(2,3,3,5)^T,
    \label{eq:FibPhiVector}
\end{equation}
One could also have used Eq. ~\eqref{eq:Nk_from_N1} to obtain the general structure of the multiplicity vector:
\begin{equation}
	\mathbf{n}^{k} = (F_{2k-1},F_{2k},F_{2k},F_{2k+1}),
    \label{eq:FibGeneralVector}
\end{equation}
where $F_n$ denotes the $n^{\rm{th}}$ Fibonacci number.

Interestingly, the dimension of the vacuum object $[\mathbf{n}^\phi]_\mathbf{1}=2$, meaning that there are two simple idempotents, $p_{\mathbf{1}}^{(11)}$ and $p_{\mathbf{1}}^{(\tau \tau)}$, that compose this central idempotent. Similar to the previous cases, $p_{\mathbf{1}}^{(11)}$ projects onto a subspace of the algebra where all the tails are trivial, while  $p_{\mathbf{1}}^{(\tau \tau)}$ projects onto a subspace where both tails are labeled by $\tau$. 

As mentioned in section \ref{subsec:Generalizing_To_LW_k}, this has implications on the $LW_2$ model as now the vacuum can exist in more than one subtype. On the one hand, the projector operator $\Pi_{\mathbf{1}}^{(11)}(p)$, associated to the first simple idempotent, corresponds to the Kirby strand and is equal to the plaquette operator $B_p$ of the Hamiltonian, such that it projects onto a ``true vacuum" where there are no charge or flux quantum numbers inside the plaquette. On the other hand, eigenstates of the projector $\Pi_{\mathbf{1}}^{(\tau\tau)}(p)$ have both tails excited (i.e. with non-trivial label). Because the operator is measuring the total quantum number, from both tails and plaquette, these two tails are together in the vacuum, thus we interpret that  $\Pi_{\mathbf{1}}^{(\tau\tau)}(p)$ projects onto a ``false vacuum".

The action of the $S$-matrix on the multiplicity vector \eqref{eq:FibPhiVector} yields
\begin{equation}
    S\mathbf{n}^\phi = \left(\left(\varphi^2\right)^2,0,0,\left(\varphi^{-2}\right)^2\right)^T,
\end{equation}
in agreement with the proposition of Eq.~\eqref{eq:Snk}. More generally, we can also verify Eq. ~\eqref{eq:Snk} using the multiplicity vector ~\eqref{eq:FibGeneralVector}.

\section{\label{sec:Conclusion} Conclusion}
In this paper, we have fully characterized the energy spectrum of a family of extended string-net models, which realize all the anyon excitations of the emergent topological order. Inspired by recent studies of extended string-net models, originally introduced in~\cite{Hu_2018}, we introduced string-net models $LW_k$ parametrized by the number $k\geq 0$ of ``tails" present at each plaquette of an input trivalent graph. For each of these models, there is a corresponding tube algebra $\mathcal{TA}_k$ whose simple modules fully describe the anyons of the emergent topological order. When $k\geq 1$, these algebras are Morita equivalent, meaning that all the extended models realize the same topological order. We highlight the importance of the $k=1$ tube algebra, called the $Q$-algebra, detailing how the dimensions of the central idempotents of $\mathcal{TA}_k$ can be obtained from the knowledge of the $Q$-algebra simple modules.

Our main results include an expression for the spectral degeneracies of extended string-net models for arbitrary input data. In addition to the previously known topological degeneracies, it also includes local, non-topological degrees of freedom which we interpret as subtypes of the anyons. These subtypes are described by the dimensions $n_A^{(k)}$ of the corresponding tube algebra’s central idempotents. From this expression, the partition function is computed, providing a tool to inspect several finite-temperature properties and their dependence on the size of the input graph.  Contrary to the zero temperature description of these extended models, which depends only on $\dc$, the form of the partition function shows that their finite-temperature description depends also on the input category $\mathcal{C}$ via $n_A^{(k)}$.

Additionally, we report that the topological orders arising in these extended string-net models are destroyed at any temperature $T>0$ in the thermodynamic limit. This is done by direct inspection of different probes, namely the specific heat, Wegner-Wilson loops and topological mutual information, which only display signs of a topologically-ordered phase at $T=0$. Away from the thermodynamic limit, some signatures of topological order are preserved, and we detail a scaling behavior of finite-size systems with temperature, which generalizes the findings by Iblisdir et al.~\cite{Iblisdir_2010,Iblisdir2009,Castelnovo2007,Nussinov2008} for the Kitaev quantum double model and the recent results for the $LW_0$ model ~\cite{Ritz-Zwilling2024_2}. In line with these references, we find that this scaling behavior is that of the 1D classical Potts model with $q$ states, with $q$ not necessarily an integer.

Importantly, we also identify a subset $\mathcal{N}$ composed of at most two fluxons: the vacuum $\mathbf{1}$, which always drives the thermodynamic limit, and, if the input UFC $\mathcal{C}$ is non-Abelian, another fluxon $B$ that characterizes the corrections to the thermodynamic limit. The identification of the anyon $B$ is non-trivial and requires knowledge about the input UFC and the tube algebra's tensors, as detailed in Appendix~\ref{appendix:SpecialSubset}.

\acknowledgements
This work was started during the master internship of A.S. in Paris and Orsay. We acknowledge the precious help of Andrej Mesaros during the internship and useful discussions with Laurens Lootens, Steve Simon and Julien Vidal. In particular, we thank Laurens for explaining to us the concrete algorithm used to decompose a tube algebra into idempotents (see Appendix~\ref{subsec:Algorithm}). A.S. also acknowledges the support of the Erasmus+: Erasmus Mundus programme of the European Union.

\appendix

\section{\label{appendix:SpecialSubset} Subset of fluxons \texorpdfstring{$\mathcal{N}$}{Lg}}
A key quantity appearing in the partition functions for the extended string-net models is the action of the modular $S$-matrix of $\dc$ on the multiplicities vectors $\mathbf{n}^{(k)}$ arising from the $\mathcal{TA}_k$ of $\mathcal{C}$. Here we present some general arguments that enable the computation of $S\mathbf{n}^Q$ in some special cases. After determining $S\mathbf{n}^Q$, one can simply use Eq.~\eqref{eq:Snk} to obtain $S\mathbf{n}^{(k)}$. This procedure also reveals the fluxons belonging to the subset $\mathcal{N}$, defined in section \ref{subsec:subset}, because they satisfy $[S\mathbf{n}^Q]_A\neq 0$. This subset contains either only the vacuum, if the input category $\mathcal{C}$ is Abelian, or the vacuum and another fluxon otherwise. Note that the subset $\mathcal{N}$ is not, in general, the same as the subset of pure fluxons $\mathcal{P}$, which is defined by $A\in \mathcal{P}$ iff $n_{A,1}=d_A$~\cite{Ritz-Zwilling2024_2}.

\subsection{\texorpdfstring{$Q$-}{Lg}algebra sectors \label{subAppendix:TAsectors}}

Tubes in the $Q$-algebra are labeled by 4 strings $\mathcal{T}^{i;j}_{r;s}$, where $r,s$ are the two tails and $i,j$ do not have open ends and wind around the periodic direction of the tube. A closed formula can be obtained for the composition of tubes. Using standard notation for the $F$-symbols (see ~\cite{LanWen, Simon2023}, for example), we find:
\begin{equation}
    \mathcal{T}_{r;s}^{i;j} \mathcal{T}_{s^{\prime};t}^{k;l}=  \sum_{m, n} \left[ \delta_{s s^{\prime}} \sqrt{\frac{d_i d_k}{d_m}} F_{\bar{k} l \bar{n}}^{i \bar{j} s} F_{\bar{k} n \bar{m}}^{\bar{r} \bar{j} j} F_{i \bar{n} m}^{t k \bar{l}} \right] \mathcal{T}_{r;t}^{m;n}.
\end{equation}

Since the open strings must match when composing tubes, one can observe that the algebra decomposes into sectors labeled by the tails $rs$. In addition, one can identify ``diagonal" sectors $rr$ (i.e. $s=r$) that are independent of each other. The decomposition of the tube algebra into these sectors greatly facilitates the search for simple idempotents and nilpotents, since idempotents necessarily belong to the ``diagonal" sectors $rr$ while nilpotents belong to the ``off-diagonal" sectors $rs, r\neq s$.

Here, however, we will be interested in the projectors $Q_r$ onto the diagonal $rr$ sectors of the $Q$-algebra. A simple way to write these projectors is to use the tubes $\mathcal{T}^{1;r}_{r;r}$, since:

\begin{equation}
    \mathcal{T}^{1;r}_{r;r} \mathcal{T}^{k;l}_{r;r} = \mathcal{T}^{k;l}_{r;r} \mathcal{T}^{1;r}_{r;r} = \mathcal{T}^{k;l}_{r;r}.
\end{equation}

Graphically, pre/post multiplication by $\mathcal{T}^{1;r}_{r;r}$ simply extends an open string $r$ of a tube in the $rr$ sector, and yields 0 if the tube is not in this sector, thus $Q_r = \mathcal{T}^{1;r}_{r;r}$

Alternatively, one can build $Q_r$ through the simple idempotents. For the $Q$-algebra, they can be labeled $p_A^{(r,a)}$ with $A$ being an irreducible block of the algebra, $r\in \mathcal{C}$ and $a\leq n_{A,s}$ \cite{Ritz_Zwilling_2024}. Naturally, $p_A^{(r,a)}$  belongs to the $rr$ sector, since it will only be composed by tubes in this sector [see Eq.~\eqref{eq:QAlgebra_Modules}]. Thus the projector $Q_r$ is given by:

\begin{equation}
    Q_r = \sum_{A} \sum_{a=1}^{n_{A,r}} p_A^{(r,a)(r,a)}
\end{equation}

\subsection{From multiplicities to tubes}

To motivate a relation between the multiplicities $n^Q_A$ and the projectors $Q_r$, let us construct another set of vectors encoding the multiplicities. Within this algebra, the internal multiplicities $n_{A,s}$ may be interpreted as the number of strings $s$ to which the anyon $A$ decomposes into (cf. discussion in section \ref{subsec:Idempotents_and_Multiplicities}). The entries of the vector $\mathbf{n}^Q$ are given by:

\begin{equation}
    [\mathbf{n}^Q]_A = \sum_{s \in \mathcal{C}} n_{A,s}.
\end{equation}

Alternatively, we can define $|\mathcal{C}|$ vectors $\mathbf{n}^Q_r, r\in \mathcal{C}$, each with entries:

\begin{equation}
    [\mathbf{n}^Q_r]_A = n_{A,r},
\end{equation}
such that $\mathbf{n}^Q = \sum_r \mathbf{n}^Q_r$. Each of these vectors $\mathbf{n}^Q_r$ is closely related to the projectors $Q_r$, which may be seen from the following relation:

\begin{equation}
    \text{Tr}\left(Q_r\right) =  \sum_{A} \sum_{a=1}^{n^Q_{A,r}}  \text{Tr} \left(p_A^{(r,a)(r,a)}\right) =\sum_{A} [\mathbf{n}^Q_r]_A.
\end{equation}

We propose that this relation can be extended by treating $Q_r$ and $\mathbf{n}^Q_r$ on the same footing. This is very helpful in computing the quantity $S\mathbf{n}^Q$ because the action of the $S$-matrix on the tubes is well-known \cite{Simon2023}. For tubes in the diagonal sectors, the open-ended strings can be glued together and the tubes become embedded on a torus. The action of the $S$-matrix on these tubes is to exchange the longitudinal and meridional directions of the torus.

Because the projector $Q_r$ is equal to the tube $\mathcal{T}_{r;r}^{1;r}$, we obtain that

\begin{equation}
    SQ_r = S \mathcal{T}^{1;r}_{r;r} = \mathcal{T}_{1;1}^{r;r}.
    \label{eq:SQ_r}
\end{equation}

Substituting $Q_r$ by $\mathbf{n}^Q_r$ and using the definition of the tube algebra's tensors in Eq.~\eqref{eq:IdempotentDef}, we arrive at:

\begin{equation}
\begin{split}
      S\mathbf{n}^Q & =  \sum_{r\in \mathcal{C}} SQ_r =  \sum_{r\in \mathcal{C}} \mathcal{T}_{1;1}^{r;r}\\
      &=\sum_{r\in \mathcal{C}}  \sum_{A\in \dc}\sum_{a,b=1}^{n_{A,r}} \left(M_A^{(1,a)(1,b)} \right)^{r}_{11r} p_A^{(1,a)(1,b)},
\end{split}
\end{equation}
and we can extract the individual components $[S\mathbf{n}^Q]_A$ of the resulting vector:

\begin{equation}
    [S\mathbf{n}^Q]_A = \sum_{r\in \mathcal{C}}\sum_{a,b} \left( M_A^{(1,a)(1,b)} \right)^{r}_{11r}.
    \label{eq:SnQ_HalfBraids}
\end{equation}

Although it is difficult to further evaluate this expression in the general case, the structure of Eq.~\eqref{eq:SnQ_HalfBraids} reveals that in order for $[S\mathbf{n}^Q]_A \neq 0$, there must be at least one simple idempotent satisfying Tr$(p_A^{(1,a)(1,a)})\geq 1$, which is the condition for $A$ to be a fluxon. In summary, we have proven that:
\be
[S\mathbf{n}^Q]_A =0 \text{ if } A\notin \mathcal{F},
\label{eq:proof1}
\ee
which means that $\mathcal{N}\subseteq \mathcal{F}$. 

In addition, for the vacuum $A=\mathbf{1}$, the tube algebra's tensors are known to be $\left( M_{\mathbf{1}}^{1,1} \right)^{r}_{11r} = d_r$, yielding Eq.~\eqref{eq:SN_1}:
\be
[S\mathbf{n}^Q]_\mathbf{1} = \sum_{r \in \mathcal{C}} d_r.
\label{eq:proof2}
\ee

We can also finish the calculation of Eq.~\eqref{eq:SQ_r} when $\mathcal{C}$ is Abelian, i.e. such that $d_r=1$, $\forall r\in \mathcal{C}$. Then we know that 
\be
\sum_{r\in \mathcal{C}} \mathcal{T}_{1;1}^{r;r} = \mathcal{D} p_\mathbf{1}^{11} = \mathcal{D} P_\mathbf{1},
\ee
see, e.g., Eq.~(A25) in~\cite{Ritz_Zwilling_2024}. This means that 
\be
[S\mathbf{n}^Q]_A = \mathcal{D} \delta_{A,\mathbf{1}},
\label{eq:proof3}
\ee
and, in that case, the subset $\mathcal{N}=\{\mathbf{1}\}$.

In addition to the three proven results Eqs.~\eqref{eq:proof1}, \eqref{eq:proof2} and \eqref{eq:proof3}, and based on our experience with several examples, we conjecture the following result. 

If $\mathcal{C}$ is non-Abelian, then $\mathcal{N}=\{\mathbf{1},B\}\subseteq \mathcal{F}$ such that:
\begin{equation}
(S\mathbf{n}^Q)_\mathbf{1} = \sum_{r\in \mathcal{C}} d_r > (S\mathbf{n}^Q)_B /d_B >0
\end{equation}

\subsection{Conclusion}
Table~\ref{table:subsets} summarizes the role played by four subsets of anyons (i.e. vacuum anyon $\{\mathbf{1}\}$, pure fluxons $\mathcal{P}$, subset of fluxons $\mathcal{N}$, and fluxons $\mathcal{F}$) in the partition function of the $LW_k$ models. In the thermodynamic limit, the partition function is dominated by the anyons $A$ with the largest $(S\mathbf{n}^{(k)})_A / d_A$. However, in a finite-size system, the anyons $A$ that satisfy $(S\mathbf{n}^{(k)})_A \neq 0$ also play special role in the partition function [see Eqs.~(\ref{eq:PartitionFunction_LW1},\ref{eq:PartitionFunction_LW0},\ref{eq:PartitionFunction_LWk})]. 
\begin{table}[h!]
\begin{center}
\begin{tabular}{|c ||c| c|} 
\hline
 & finite size & thermodynamic limit \\ 
\hline\hline
$LW_0$ & $\mathcal{F}$ i.e.  $n^Q_{A,1}\neq 0$ & $\mathcal{P}$ i.e.  $n^Q_{A,1} = d_A$ \\ 
\hline
$LW_{k\geq 1}$ & $\mathcal{N}$ i.e.  $(S \mathbf{n}^{Q})_A\neq 0$ & $\{\mathbf{1}\}$ \\ 
\hline
\end{tabular}
\caption{Four subsets of anyons ($\{\mathbf{1}\}$, $\mathcal{P}$, $\mathcal{N}$ and $\mathcal{F}$) that play a special role in the partition function of the $LW_k$ models.}
\label{table:subsets}
\end{center}
\end{table}

\section{\label{appendix:TAdecomp} Idempotent decomposition of tube algebras}

The Artin-Wedderburn theorem states that any semisimple algebra $\mathcal{A}$ has an unique decomposition into a sum of simple ideals $\mathcal{A}=\mathcal{A}_1\oplus\mathcal{A}_2\oplus\dots\oplus\mathcal{A}_c$, and each simple ideal is isomorphic to a full matrix algebra~\cite{Drozd1994}. Constructive algorithms to the Artin-Wedderburn theorem are well-known in the mathematical literature (see for example~\cite{WedderburnDecomp,Ivanyos_1999,EfficientDecomposition}). They are mostly built to decompose algebras over finite or rational number fields (i.e. extensions of $\mathbb{Q}$), which do not include $\mathbb{C}^{*}$, the base field of the tube algebras introduced in this work. Nonetheless the algorithm proposed in \cite{WedderburnDecomp} (also reviewed in Appendix E of \cite{Bultinck_2017}) has already been successfully used in \cite{Bultinck_2017} to decompose some $Q$-algebras.  Here, this algorithm, closely following ~\cite{WedderburnDecomp}, was also used to decompose the $\phi$-algebra for some input categories. However, one should be careful to note that, unlike the algorithms presented in the mathematical literature, the one presented below is not completely analytical, as some of the steps involve numerically finding eigenvectors of randomly generated matrices.

\subsection{\label{subsec:Algorithm} Idempotent decomposition of a semisimple algebra}

The algorithm takes as input the structure constants $f_{ij}^k$ of the algebra $\mathcal{A}$. For us, this algebra is a tube algebra and its basis elements are tubes that are represented by a canonical vector space basis $\{b_1,\dots,b_r\}$, such that $b_ib_j=\sum_k f_{ij}^k b_k$, where $r$ is the dimension of the tube algebra $\mathcal{A}$. It will output the simple idempotents defined in Eq.~\eqref{eq:CentralIdemp_def} as linear combinations of the basis elements.

Firstly, the problem is reduced to the decomposition of the algebra center  $\mathfrak{C}$, composed of elements $x$ that satisfy $xb_i=b_ix, \forall b_i\in \mathcal{A}$. This center of the tube algebra should not be confused with the Drinfeld center, which is composed by the central idempotents. Indeed, a semisimple algebra and its center algebra share the same decomposition~\cite{Ivanyos_1999}:
\begin{equation}
    \mathcal{A} \cong \bigoplus_{i=1}^c\mathcal{A}_i \Longrightarrow
            \mathfrak{C} \cong \bigoplus_{i=1}^c\mathfrak{C}_i.
    \label{eq:TAtoCenter}
    \end{equation}
    
The quantity $c$ simultaneously stands for the number of simple matrix algebras that arise from the Artin-Wedderburn decomposition of $\mathcal{A}$ [first equality in Eq.~\eqref{eq:TAtoCenter}], defines the dimension of the algebra $\mathfrak{C}$ [second equality in Eq.~\eqref{eq:TAtoCenter}] and will later describe the order of the Drinfeld center. Furthermore, it stands that $\mathcal{A}_i =\mathcal{A} \, \mathfrak{C}_i$, where $\mathcal{A} \, \mathfrak{C}_i$ denotes the algebra whose elements are of the form $bx_i$ for $b\in \mathcal{A}$ and $x_i\in \mathfrak{C}_i$.
The elements $x \in \mathfrak{C}$ can be computed by solving the system of equations spanned by $xb_i-b_ix=0$ for each basis element $b_i$. Thus we can represent these elements by the null space of the following $r^2\times r$ matrix:
\begin{equation}
     Z_{(i-1)r+k,j} = f^k_{ij} - f^k_{ji}, \,\, \forall i,j,k = {1,\dots,r}
\end{equation}

Let $\{x_1,\dots,x_c\}$ be a canonical basis for the nullspace of $Z$, corresponding to the basis elements of the commutative algebra $\mathfrak{C}$. Note that each element $x_i$ is an element of $\mathcal{A}$, and has $r$ components in the basis $\{b_j\}$. The structure constants $d_{ij}^k$ that define the multiplication $x_ix_j = \sum_k d_{ij}^k x_k$   in $\mathfrak{C}$ can be obtained by determining the row reduced echelon form (RREF)  of the matrix
\begin{equation*}
    \left[ x_1 \quad x_2 \quad \dots \quad x_c \quad v_{ij}\right],
\end{equation*}
 where the first $c$ columns are the basis vectors, and $v_{ij}=x_ix_j$ is the column vector representation of the product $x_ix_j$ in the basis $\{b_j\}$, calculated using the structure constants $f_{ij}^k$. The RREF of the above matrix yields the structure constants $d_{ij}^k$ in the last column.

These structure constants can be thought of as matrices $d_i$, representing the multiplication/composition of elements in the algebra:
\begin{equation}
    x_ix_j = \sum_{k}^c d_{ij}^k x_k \equiv \sum_{k}^c [d_i]_j^k x_k.
\end{equation}

Each $d_i$ is a $c^2$ dimensional matrix, such that it can be thought of as a particular sum of idempotents or nilpotents. 

Next, we consider the following matrix
\begin{equation}
    X = \sum_i^c \zeta_i d_i,
\end{equation}
where $\zeta_i$ are random complex numbers picked uniformly from the complex unit square. The $c$ eigenvectors of this matrix are the $c$ central idempotents $e_A$ of the center algebra $\mathfrak{C}$, up to a normalization factor~\cite{Lootens2024}. We label the central idempotents $e_A$ with a capital latin letter (such as $A$) to make connection to the notation used in the core of the article.

After normalizing each central idempotent, we can rewrite them on the basis $\{b_k\}$ of $\mathcal{A}$. Let $(e_A)_j$ denote the components of the central idempotent $e_A$ on the basis $\{x_j\}$, and $(x_j)_k$ the components of the basis vectors $x_j$ on the basis $\{b_k\}$. Then the central idempotents $P_A$ of the original algebra $\mathcal{A}$ are given by:
\begin{equation}
    P_A = \sum_{j=1}^c \sum_{k=1}^r \left(e_A\right)_j \left(x_j\right)_k b_k.
    \label{eq:CenterToFull}
\end{equation}

The dimensions $n_A$ of each algebra, that correspond to the internal multiplicities, are obtained by computing the dimension of the vector space spanned by each $P_A$.

\subsection{Decomposition of Fib \texorpdfstring{{$\phi$}}{Lg}-algebra}

Tubes in the $\phi$-algebra have two tails on each side of the tube, as depicted in Fig.~\ref{fig:Tube2}. There are a total of 47 tubes for the Fib UFC and using the notation introduced in Fig.~\ref{fig:Tube2} to label them, the structure constants can be obtained as follows: 
\begin{widetext}
    \begin{equation}
    \mathcal{T}_{ab;cd}^{e;fgh}\mathcal{T}_{ij;kl}^{m;nop} = \delta_{c,i}\delta_{d,j} \left[ \sum_{wxyz} \sqrt{\frac{d_ed_m}{d_w}\frac{d_gd_o}{d_y}} F^{\bar{x}gn}_{o\bar{k}y}F^{\bar{g}bh}_{z\bar{o}y} \sum_{p'h'} 
    F^{\bar{o}dp}_{m\bar{l}p'}
    F^{\bar{h}ed}_{m\bar{p'}w}
    F^{\bar{e}af}_{g\bar{c}f'}
    F^{\bar{e}f'\bar{c}}_{n\bar{m}w}
    F^{ag\bar{f'}}_{n\bar{w}x}
    F^{\bar{h}w\bar p'}_{\bar{l}\bar{o}z} \right] \, \mathcal{T}_{ab;kl}^{w;xyz}.
    \end{equation}
\end{widetext}
Inserting these structure constants in the decomposition algorithm, we are able to find 4 central idempotents as combinations of tubes:
\begin{widetext}

\begin{equation*}
\begin{split}
&\begin{split}
P^{\mathbf{1}} = \frac{1}{\sqrt{5}} \bigg(\,
\frac{1}{\varphi}\mathcal{T}_{00;00}^{0;000} +\mathcal{T}_{00;00}^{\tau;\tau\tau\tau} 
+ \frac{1}{\varphi^3}\mathcal{T}_{\tau\tau;\tau\tau}^{0;\tau0\tau}
&+\frac{1}{\varphi^{-5/2}}\mathcal{T}_{\tau\tau;\tau\tau}^{0;\tau\tau\tau}
+\frac{1}{\varphi^3}\mathcal{T}_{\tau\tau;\tau\tau}^{\tau;0\tau0}
+\frac{1}{\varphi^{-5/2}}\mathcal{T}_{\tau\tau;\tau\tau}^{\tau;0\tau\tau} \\
&+\frac{1}{\varphi^{-5/2}}\mathcal{T}_{\tau\tau;\tau\tau}^{\tau;\tau0\tau}
+\frac{1}{\varphi^{-5/2}}\mathcal{T}_{\tau\tau;\tau\tau}^{\tau;\tau\tau0}
-\frac{1}{\varphi^3}\mathcal{T}_{\tau\tau;\tau\tau}^{\tau;\tau\tau\tau}
   \, \bigg)
\end{split}
\\
\\
&\begin{split}
&P^{\tau 1 } = \frac{1}{\sqrt{5}} \bigg(\,
\frac{1}{\varphi}\mathcal{T}_{0\tau;0\tau}^{0;00\tau}
+\frac{e^{\frac{4\pi i }{5}}}{\varphi}\mathcal{T}_{0\tau;0\tau}^{\tau;\tau\tau0}
+\frac{e^{-\frac{3\pi i }{5}}}{\sqrt{\varphi}}\mathcal{T}_{0\tau;0\tau}^{\tau;\tau\tau\tau}
+\frac{1}{\varphi}\mathcal{T}_{\tau0;\tau0}^{0;\tau00}
+\frac{e^{\frac{4\pi i }{5}}}{\varphi}\mathcal{T}_{\tau0;\tau0}^{\tau;0\tau\tau}
+\frac{e^{-\frac{3\pi i }{5}}}{\sqrt{\varphi}}\mathcal{T}_{\tau0;\tau0}^{\tau;\tau\tau\tau}\\
&+ \frac{1}{\varphi^2}\mathcal{T}_{\tau\tau;\tau\tau}^{0;\tau0\tau}
-\frac{1}{\varphi^{5/2}}\mathcal{T}_{\tau\tau;\tau\tau}^{0;\tau\tau\tau}
+\frac{e^{-\frac{3\pi i }{5}}}{\varphi^2}\mathcal{T}_{\tau\tau;\tau\tau}^{\tau;0\tau0}
+\frac{e^{\frac{2\pi i }{5}}}{\varphi^{5/2}}\mathcal{T}_{\tau\tau;\tau\tau}^{\tau;0\tau\tau}
-\frac{1}{\varphi^{5/2}}\mathcal{T}_{\tau\tau;\tau\tau}^{\tau;\tau0\tau}
+\frac{e^{\frac{2\pi i }{5}}}{\varphi^{5/2}}\mathcal{T}_{\tau\tau;\tau\tau}^{\tau;\tau\tau0}
+ \frac{5^{1/4}e^{\frac{7\pi i }{10}}}{\varphi^{5/2}}\mathcal{T}_{\tau\tau;\tau\tau}^{\tau;\tau\tau\tau}
   \, \bigg)
\end{split}
\end{split}
\end{equation*}
\\
\\
\begin{equation*}
\begin{split}
&\begin{split}
&P^{1 \overline{\tau}} = \frac{1}{\sqrt{5}} \bigg(\,
\frac{1}{\varphi}\mathcal{T}_{0\tau;0\tau}^{0;00\tau}
+\frac{e^{-\frac{4\pi i }{5}}}{\varphi}\mathcal{T}_{0\tau;0\tau}^{\tau;\tau\tau0}
+\frac{e^{\frac{3\pi i }{5}}}{\sqrt{\varphi}}\mathcal{T}_{0\tau;0\tau}^{\tau;\tau\tau\tau}
+\frac{1}{\varphi}\mathcal{T}_{\tau0;\tau0}^{0;\tau00}
+\frac{e^{-\frac{4\pi i }{5}}}{\varphi}\mathcal{T}_{\tau0;\tau0}^{\tau;0\tau\tau}
+\frac{e^{\frac{3\pi i }{5}}}{\sqrt{\varphi}}\mathcal{T}_{\tau0;\tau0}^{\tau;\tau\tau\tau}\\
&+ \frac{1}{\varphi^2}\mathcal{T}_{\tau\tau;\tau\tau}^{0;\tau0\tau}
-\frac{1}{\varphi^{5/2}}\mathcal{T}_{\tau\tau;\tau\tau}^{0;\tau\tau\tau}
+\frac{e^{\frac{3\pi i }{5}}}{\varphi^2}\mathcal{T}_{\tau\tau;\tau\tau}^{\tau;0\tau0}
+\frac{e^{-\frac{2\pi i }{5}}}{\varphi^{5/2}}\mathcal{T}_{\tau\tau;\tau\tau}^{\tau;0\tau\tau}
-\frac{1}{\varphi^{5/2}}\mathcal{T}_{\tau\tau;\tau\tau}^{\tau;\tau0\tau}
+\frac{e^{-\frac{2\pi i }{5}}}{\varphi^{5/2}}\mathcal{T}_{\tau\tau;\tau\tau}^{\tau;\tau\tau0}
+ \frac{5^{1/4}e^{-\frac{7\pi i }{10}}}{\varphi^{5/2}}\mathcal{T}_{\tau\tau;\tau\tau}^{\tau;\tau\tau\tau}
   \, \bigg)
\end{split}
\\
\\
&\begin{split}
P^{\tau \overline{\tau} } =& \frac{1}{\sqrt{5}} \bigg(\,
\varphi\mathcal{T}_{00;00}^{0;000} 
-\mathcal{T}_{00;00}^{\tau;\tau\tau\tau} 
+\mathcal{T}_{0\tau;0\tau}^{0;00\tau}
+\mathcal{T}_{0\tau;0\tau}^{\tau;\tau\tau0}
+\frac{1}{\varphi^{3/2}}\mathcal{T}_{0\tau;0\tau}^{\tau;\tau\tau\tau}
+\mathcal{T}_{\tau0;\tau0}^{0;\tau00}
+\mathcal{T}_{\tau0;\tau0}^{\tau;0\tau\tau}
+\frac{1}{\varphi^{3/2}}\mathcal{T}_{\tau0;\tau0}^{\tau;\tau\tau\tau}\\
&+ \frac{2}{\varphi}\mathcal{T}_{\tau\tau;\tau\tau}^{0;\tau0\tau}
+\frac{1}{\varphi^{5/2}}\mathcal{T}_{\tau\tau;\tau\tau}^{0;\tau\tau\tau}
-\frac{1}{\varphi^{3/2}}\mathcal{T}_{\tau\tau;\tau\tau}^{\tau;0\tau\tau}
+\frac{1}{\varphi^{5/2}}\mathcal{T}_{\tau\tau;\tau\tau}^{\tau;\tau0\tau}
-\frac{1}{\varphi^{3/2}}\mathcal{T}_{\tau\tau;\tau\tau}^{\tau;\tau\tau0}
+ \frac{2}{\varphi^{2}}\mathcal{T}_{\tau\tau;\tau\tau}^{\tau;\tau\tau\tau}
   \, \bigg) .
\end{split}
\end{split}
\end{equation*}

\end{widetext}


%

\end{document}